\documentclass{emulateapj}

\usepackage{epstopdf}

\def\squig{\sim\!\!}

\def\spose#1{\hbox to 0pt{#1\hss}}
\def\simlt{\mathrel{\spose{\lower 3pt\hbox{$\mathchar"218$}}
     \raise 2.0pt\hbox{$\mathchar"13C$}}}
\def\simgt{\mathrel{\spose{\lower 3pt\hbox{$\mathchar"218$}}
     \raise 2.0pt\hbox{$\mathchar"13E$}}}

\shorttitle{M31 Spheroid Kinematics}
\shortauthors{Dorman et al.}

\slugcomment{Accepted to ApJ, 16 April 2012}

\begin{document}


%--------------------------------------------------------
%  TITLE & AUTHOR LIST
%--------------------------------------------------------
\title{The SPLASH Survey:\\ Kinematics of Andromeda's Inner Spheroid}
\author{Claire~E.\ Dorman\altaffilmark{1}, 
Puragra\ Guhathakurta\altaffilmark{1}, 
Mark~A.\ Fardal\altaffilmark{2},
Dustin Lang\altaffilmark{3},
Marla~C.\ Geha\altaffilmark{4},
Kirsten~M.\ Howley\altaffilmark{5}, 
Jason~S.\ Kalirai\altaffilmark{6, 7}, 
James~S.\ Bullock\altaffilmark{8},
Jean-Charles\ Cuillandre\altaffilmark{9},
Julianne~J.\ Dalcanton\altaffilmark{10},
Karoline~M.\ Gilbert\altaffilmark{10,11}, 
Anil~C. Seth\altaffilmark{12}, 
Erik~J. Tollerud\altaffilmark{8},
Benjamin~F. Williams\altaffilmark{9},
Basilio\ Yniguez\altaffilmark{8}
}

\altaffiltext{1}{UCO/Lick Observatory, University of California
    Santa Cruz, 1156 High Street, Santa Cruz, CA~95064; {\tt [cdorman,
      raja]@ucolick.org}}
\altaffiltext{2}{Department of Astronomy, University of
  Massachussetts, Amherst, MA~01003; {\tt fardal@astro.umass.edu}}
\altaffiltext{3}{Department of Astrophysical Sciences, Princeton
  University, Princeton, NJ 08544; {\tt dstn@astro.princeton.edu}}
\altaffiltext{4}{Department of Astronomy, Yale University, New Haven,
  CT~06510; {\tt marla.geha@yale.edu}}
\altaffiltext{5}{Lawrence Livermore National Laboratory, P.O. Box 808, 
Livermore, CA~94551; {\tt howley1@llnl.gov}}
\altaffiltext{6}{Space Telescope Science Institute, 3700 San Martin
  Drive, Baltimore, MD~21218; {\tt jkalirai@stsci.edu}}      
\altaffiltext{7}{Center for Astrophysical Sciences, Johns Hopkins
  University, Baltimore, MD~21218}
\altaffiltext{8}{Department of Physics \& Astronomy, University of
  California, Irvine, 4129 Frederick Reines Hall, Irvine, CA~92697;
  {\tt [bullock, etolleru, byniguez]@uci.edu}}
\altaffiltext{9}{Canada-France-Hawaii Telescope, 65-1238 Mamalahoa
  Hwy, Kamuela, HI~96743; {\tt jcc@cfht.hawaii.edu}}
\altaffiltext{10}{Department of Astronomy, University of Washington,
  Box 351580, Seattle, WA 98195; {\tt [jd, kgilbert,
    ben]@astro.washington.edu}}
\altaffiltext{11}{Hubble Fellow}
\altaffiltext{12}{Department of Physics \& Astronomy, University of
  Utah, Salt Lake City, UT 84112; {\tt aseth@astro.utah.edu}}

\altaffiltext{13}{The W.~M.~Keck
  Observatory is operated as a scientific partnership among the
  California Institute of Technology, the University of California,
  and NASA. The Observatory was made possible by the generous
  financial support of the W.~M.~Keck Foundation.}
%--------------------------------------------------------
 %  ABSTRACT
%--------------------------------------------------------
\begin{abstract}
The combination of large size, high stellar density, high metallicity,
and S\'ersic surface brightness profile of the spheroidal component of
the Andromeda galaxy (M31) within $R_{\rm proj}\sim 20$~kpc
suggest that it is unlike any subcomponent of the Milky
Way. In this work we capitalize on our proximity to and external view
of M31 to probe the kinematical properties of this ``inner spheroid.''
We employ a Markov chain Monte Carlo (MCMC) analysis of
resolved stellar kinematics from Keck\altaffilmark{13}/DEIMOS spectra
of  $5651$ red giant branch stars to disentangle M31's inner spheroid
from its stellar disk. We measure the mean velocity and dispersion of
the spheroid in each of five spatial bins after accounting for a
locally cold stellar disk as well as the Giant Southern Stream and
associated tidal debris. For the first time,
we detect significant spheroid rotation ($v_{\rm rot}\sim 50~\rm
km~s^{-1}$) beyond $R_{\rm proj}\sim 5$~kpc. The velocity dispersion
decreases from about $140\rm~km~s^{-1}$ at $R_{\rm proj}=7$~kpc to
$120\rm~km~s^{-1}$ at $R_{\rm   proj}=14$~kpc, consistent to $2\sigma$
with existing measurements and models.  We calculate the probability that a
given star is a member of the spheroid and find that the spheroid has
a significant presence throughout the spatial extent of our
sample. Lastly, we show that the flattening of the spheroid is due to
velocity anisotropy in addition to rotation. Though
this suggests that the inner spheroid of M31 more closely resembles an
elliptical galaxy than a typical spiral galaxy bulge, it should be
cautioned that our measurements are much farther out ($2-14~r_{\rm
  eff}$) than for the comparison samples.

\end{abstract}
 
 %-------------------------------------------------------
 %  KEYWORDS
 %-------------------------------------------------------
%--------------------------------------------------------
 %  1. INTRODUCTION
 %-------------------------------------------------------
\section{INTRODUCTION}\label{intro_sec}

The elegant progression of the Hubble sequence from 
ellipticals to spirals demonstrates that galaxy morphology can
be described in large part based on the relative importance of spheroid and disk
subcomponents. While it is now clear that the
simple 
evolutionary path from elliptical ``early-type" to disk-dominated
``late-type" galaxies that \citet{hubble} originally proposed
is incorrect, the physical origin of the Hubble sequence and the
formation of and relationship between the different structural
subcomponents remain subjects of vigorous research.
% XXX this may be a good place for high z galaxy morphological studies
%Grogin+ 2011

The central spheroids of spiral
galaxies fall into two categories, which can be
explained by distinct formation mechanisms as reviewed by
\citet{kor04}.  Classical
bulges, which are typically described as elliptical galaxy analogs with random
stellar velocity distributions, large velocity dispersions, and $r^{1/4}$ de
Vaucouleurs surface brightness profiles, are likely formed through violent
merger/accretion events. Pseudobulges, which are
more flattened, have more ordered kinematics, and have roughly exponential
brightness cutoffs (or, more generally, S\'ersic profiles with low
$n_{\rm S\acute{e}rsic}$ values) are likely formed through secular
heating of the disk. More detailed observations, yielding constraints on the
structure and dynamics of bulges, will lead to a clearer understanding of
possible formation scenarios. 

Any study of the inner regions of a galaxy is complicated by the
presence of several spatially overlapping structural subcomponents, such as the
disk, spheroid, and halo. Deconvolving these
subcomponents to reveal the behavior of a single one is difficult. Traditionally,
codes such as GALFIT \citep{pen02} or GIM2D \citep{sim02} are employed
to fit galactic integrated light profiles with the sum of a S\'ersic
bulge and exponential disk \citep[e.g.][]{cou96, cou11}. This
technique is the only
possible method for characterizing the structure of distant galaxies,
but it suffers from strong assumptions about the
characteristic light profiles of bulges and disks. In addition,
degeneracy in the best-fit derived parameters can cloud interpretation
of the results.

Resolved stellar kinematics offer a complementary approach to structural
deconvolution of the nearest galaxies. Instead of assuming specific
surface brightness profiles of disks and spheroids, one must only
make the geometrical argument that a stable disk -- a thin, flat
structure -- is kinematically colder (has a higher $v_{\rm
  rot}/\sigma_v$) than a stable spheroid. Separate components can then
be identified and characterized by their distinct stellar velocity
distributions. The proximity of
Andromeda (M31) at about 785 kpc \citep[e.g.,][]{mcc05} renders it the
only large spiral galaxy other than the Milky Way (MW) where detailed
photometric {\em and} kinematical observations are possible with
current observing facilities. 

We use resolved stellar kinematics to study M31's kinematically hot
``inner spheroid'' at projected radii of 2--20 kpc. Any description
of this region -- or that in the the intermediate region of any large
galaxy -- is necessarily complex; the literature is full of vocabulary
such as ``bulge,'' ``spheroid,'' ``inner spheroid,'' ``outer
spheroid,'' ``disk,'' ``thin disk,'' ``thick disk,'' ``extended
disk,'' and so on. There is not yet a consensus on the best
combination of these nouns to represent M31. For the purposes of this
paper, we use the word ``spheroid'' to describe a kinematically hot
component: some combination of bulge, halo, and/or any other
spheroidal component. Likewise, we refer to the kinematically colder
population as the ``disk,'' where this term includes any distinct disk
components that may be present, such as the thin, thick or extended
disks.  

Despite the possibility of multiple components, the disk is
likely to be locally kinematically cold. \citet{col11} claim that M31's stellar
disk at $r_{\rm  proj}\sim$10--40~kpc consists of a cold thin disk and a warm thick
disk, as is the case for the MW.  Given that most of our fields are
closer to M31's center than the innermost field of \citet{col11},
and given their finding that the thin disk has twice the density of
the thick disk and a shorter radial scale length, we expect the cold thin disk
to dominate the stellar disk population in our fields. Similarly,
\citet{iba05} suggest that no more than about 10\% of the total disk
luminosity may be due to an extended disk component which also lags
the cold disk. Though we do not know a priori the relative
contributions of the thin, thick and extended disks, in \S\,\ref{ssec_cold}
we show that our
assumption of a dominant thin component is justified for the purposes
of measuring the kinematical parameters of the inner spheroid. 

Unlike the case of its stellar disk, M31's inner spheroid has no
analog in the MW and is therefore of great interest. 
The spheroidal system at these radii in the MW is
relatively metal-poor
\citep[$\langle Fe/H \rangle \sim -1.6$;][]{car07}, is composed
entirely of old stars, and has a power-law spatial density profile of
the form $r_{\rm deproj}^{-3}$, which corresponds to an $r^{-2}$
power-law surface brightness profile. Models such as those proposed by
\citet{bul05} and
\citet{zol10} suggest that the MW halo represents a population of
accreted dwarf satellite galaxies. In
contrast, the inner spheroid in M31 more closely resembles a bulge
than a halo. It is more metal-enhanced than the MW halo, with [Fe/H] $\sim -0.7$
\citep{kal06a}, and has a S\'ersic surface brightness profile with
$n_{\rm S\acute{e}rsic}\sim2$--4
\citep{pri94, guh05, cou11}. In addition, the stellar population of
M31's spheroid is
younger than that of the MW inner halo on average, with $40\%$ of the stars
younger than $10$ Gyr \citep{bro06}, and the stellar density is also
significantly higher than that at an equivalent location in the MW \citep{rei98}. 
 
The inner spheroid straddles territory between two well-studied
components of the spheroid: the classical
and boxy bulges interior to $\sim1~\rm kpc$
\citep{ath06,bea07,cou11}, and the outer  halo which dominates past
$R_{\rm proj} \sim 30 \rm ~kpc$  \citep[e.g.][]{guh05, irw05, iba07}. In the central
kpc, where the density is too high for resolved stellar population
spectroscopy, \citet{sag10} analyzed integrated-light kinematics to reveal a
bulge rotation speed of $70~\rm km~s^{-1}$ and a velocity dispersion
of $140~\rm km~s^{-1}$ at $R_{\rm proj}=1.1 ~\rm kpc$ on the major axis. However,
they cautioned that this measurement is contaminated by the
kinematically cold disk which may contribute nearly a third of the light
at this radius. 

Farther out in M31's halo, kinematical surveys of the resolved stellar
population using the Keck/DEIMOS multiobject spectrograph have mapped the
cold substructure, as well as the underlying smooth virialized population,
out to $R_{\rm proj}>150$~kpc \citep{guh05, cha06,
kal06a, gil07, gil09}. \citet{cha06} compiled kinematics of $\sim 1200$
red giant branch (RGB) halo stars in scattered fields between $R_{\rm proj}=$8 and
70~kpc. Using a windowing technique to eliminate stars whose velocities were
consistent with that of the disk, they found that the
velocity dispersion of the remaining population decreased radially
outwards: $\sigma_v(R_{\rm proj})=(152-0.9~R_{\rm proj} \rm ~kpc^{-1})~km~s^{-1}$. Subsequently, as
part of the Spectroscopic and Panchromatic Landscape of Andromeda's
Stellar Halo (SPLASH) survey, \citet{gil07} fit a double Gaussian
profile to the velocity distribution of RGB stars in a large
contiguous region along the southeastern minor axis of the galaxy and
measured a constant velocity dispersion of $128.9~\rm km~s^{-1}$ between $R_{\rm
 proj}=10$ and $30 \rm~kpc$.

In recent years, the focus of SPLASH has migrated inwards, first
to target the dwarf galaxies Andromeda I and Andromeda X \citep{tol11}, NGC~205
\citep{geh06, how08} and M32 \citep{how12}, and now towards the disk-
and bulge-dominated inner regions of M31. The majority of the data
analyzed in the present paper come from the most crowded area
targeted to date: a large contiguous disk-dominated area on the NE
major axis with $R_{\rm proj}=2$--19~kpc. This area was selected to overlap
the coverage of the Panchromatic Hubble Andromeda Treasury (PHAT) survey, a
five-year Hubble Space Telescope (HST) MultiCycle Treasury (MCT) program that began in 2010
\citep{dal12}.

The disk and spheroid share the inner regions of the galaxy with
remnants of tidally disrupted galaxies. The dominant features in star-count maps of the 2--20~kpc
region are the Giant Southern Stream (GSS; the remnant of a tidally
stripped satellite galaxy) and the  shelves (sharp edges in stellar
density) created by it \citep{iba01,far07}. Both the GSS and a
``secondary stream,''  which is cospatial with the GSS but separated by
$100~\rm km~s^{-1}$ in velocity, have been kinematically
detected in multiple fields south of M31 \citep{kal06b,
  gil09}. \citet{far07} identified the northern extension of the
stream in the \citet{cha06} and \citet{iba05} sample of RGBs and in the planetary
nebulae of \citet{mer06}. 

There are two principal challenges to a resolved
stellar population kinematical study of the crowded inner spheroid of M31. First,
we must select intended stellar targets whose spectra
are least likely to be contaminated by close (in projection) stellar neighbors. Second,
we must disentangle the stellar disk from the spheroid
population we wish to characterize. This is especially important in
disk-dominated fields, where fewer than $20\%$ of the stars
may belong to the spheroid. Note that though a large part of this paper will focus on accounting
for the disk contribution, our analysis method is designed to
elucidate the nature of the inner spheroid rather than the disk. We
do not attempt to make a statement here about the rotation curve of
the stellar disk or the presence of a thick disk. We plan to
analyze these components in a future paper. 

This paper is organized as follows. In \S\,\ref{sec_observations} we
explain our target selection techniques,  spectroscopic observations
and radial velocity extraction. In \S\,\ref{sec_analysis} we describe
our method for isolating and characterizing the spheroid velocity
distribution in each of five spatial bins. In \S\,\ref{sec_results} we
discuss the implications of our results; finally, we summarize our
findings in \S\,\ref{sec_summary}. 

%--------------------------------------------------------
% 2. THREE SETS OF OBSERVATIONS
%--------------------------------------------------------
\section{OBSERVATIONS}\label{sec_observations}

Our data set for this project is a compilation of three sets of RGB
spectra, two of which are presented here for the first time. A detailed technical description 
of the spectroscopic slitmask design and data reduction is given in \citet{how12}. In this 
section, we describe the target selection criteria for the different data sets and give an 
overview of the data acquisition and reduction methods common to all
the observations. 

In \S\,\ref{ssec_data}, we outline the three data sets used in this
paper. 
In  \S\,\ref{ssec_catalogs}, we describe the source catalogs from
which we select our spectroscopic targets.
In \S\,\ref{ssec_target}, we explain our target selection criteria.
In \S\,\ref{ssec_multi}, we provide the observing details.
In  \S\,\ref{ssec_mreduce}, we give a rundown of the data reduction process.  
In \S\,\ref{ssec_cc} and  \S\,\ref{ssec_quality}, we measure velocities of individual stars 
and determine the quality of those measurements, respectively.
Finally, in \S\,\ref{ssec_serendip}, we discuss the detection and velocity measurement
 of serendipitously detected stars.

%FIGURE 1
\begin{figure*}[!]
\begin{center}
\scalebox{1}{\includegraphics[trim=40 50 20 280, clip =
  true]{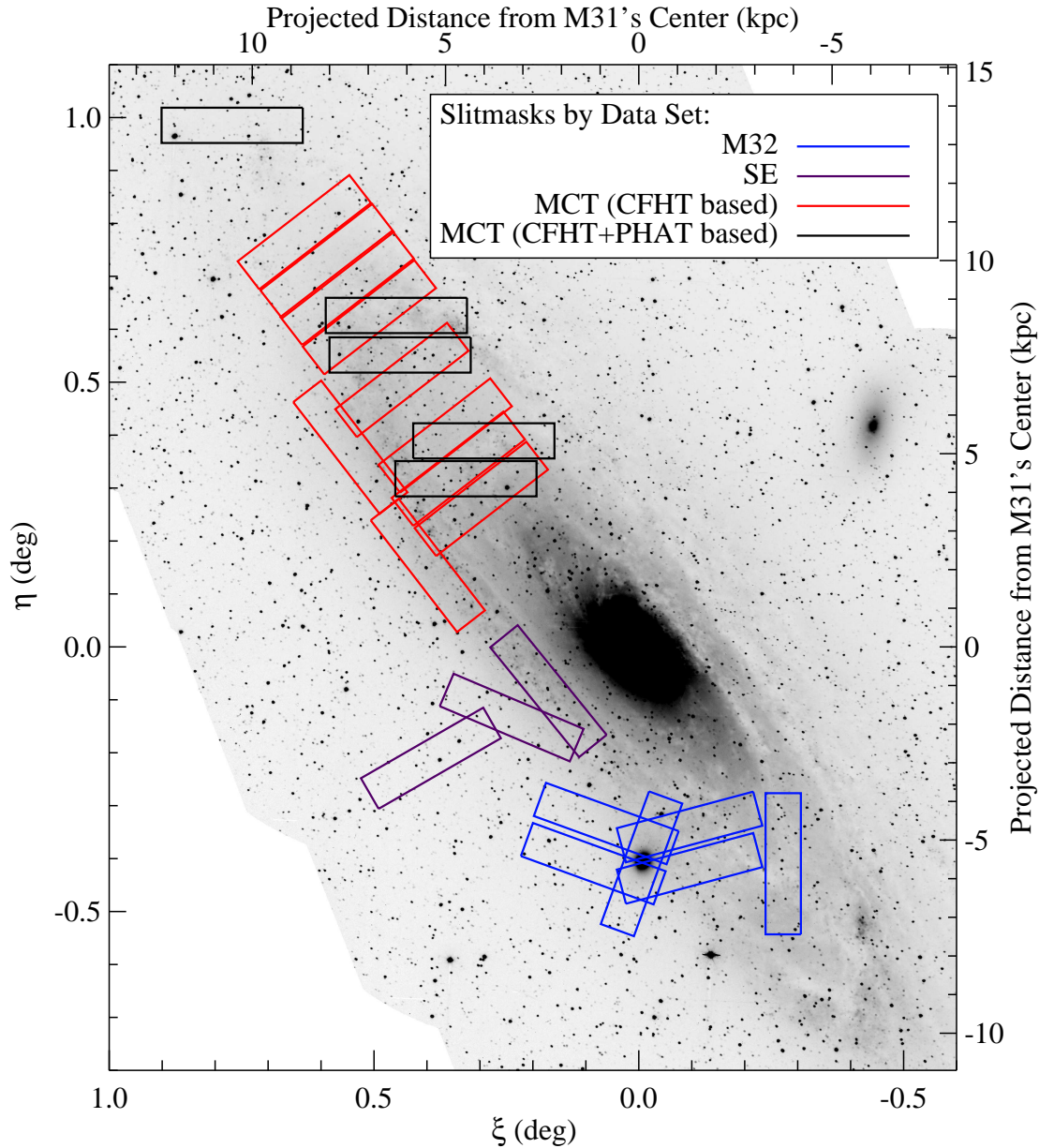}}
\centering
\caption{Twenty-four Keck/DEIMOS multiobject slitmasks overlaid on the
  \citet{cho02} KPNO Burrell  Schmidt $B$-band mosaic image of
  M31. Colored rectangles outline the  slitmasks from the different
  data sets: SE (violet), M32 (blue), and MCT (red and black,
  respectively, for slitmasks for which target selection was based on
  only the ground-based CFHT/MegaCam photometry/astrometry catalog
  versus slitmasks for which target selection was based on a
  combination of the PHAT and CFHT/MegaCam photometry/astrometry catalogs).}
\label{fig_galex}
\end{center}
\end{figure*}
\subsection{Data Sets}\label{ssec_data}
The spatial coverage of our three data sets is shown in Figure~\ref{fig_galex}. Each
rectangular outline represents a Keck/DEIMOS slitmask,
which covers approximately $16' \times 4'$ and yields  200--270 useful
spectra. (The actual footprint of a DEIMOS slitmask is not perfectly
rectangular.) The SE data sets consists of three slitmasks oriented along
the eastern minor axis of M31 (violet in Figure~\ref{fig_galex}). The
M32 data set includes the five slitmasks covering the compact elliptical
galaxy M32 directly south of M31 and one slitmask on the SW major axis
(blue in Figure~\ref{fig_galex}). The  SE and M32 data sets were
observed during the 2007 and 2008 fall seasons.

Our newest data set, from October 2010, covers a portion of M31's
northeastern major axis spanning the projected radial range
$0.33^{\circ}$--$1.38^{\circ}$  or 2--19~kpc from the center of
M31 in the plane of the disk (red and black in
Figure~\ref{fig_galex}). Five of these 15 slitmasks (black) were
chosen to maximally overlap the regions of existing  photometry from
the first year of the PHAT program.

The pre-imaging and 
reduction processes are identical for all three data sets. The primary
difference in data acquisition is in the target selection: isolated sources are
hand-selected for the SE and M32 slitmasks from a single monochromatic
ground-based source catalog \citep{how12}, while we use an 
automated series of statistical techniques as well as limited 
color information from the PHAT survey to choose targets for the MCT slitmasks.

%TABLE 1
\begin{deluxetable*}{llllrccccc}
\tabletypesize{\scriptsize}
\tablecaption{Keck/DEIMOS Multiobject Slitmask Exposures}
\tablewidth{0 pt}
\tablehead{
\colhead{Mask}&
\colhead{Observation}&
\colhead{$\alpha$ [J2000]}&
\colhead{$\delta$ [J2000]}&
\colhead{P.A.}&
\colhead{$t_{\rm{exp}}$}&
\colhead{Seeing}&
\colhead{No. of}&
\colhead{No. of Usable}&
\colhead{No. of Usable}\\
\colhead{Name}&
\colhead{Date (UT)}&
\colhead{(h:m:s)}&
\colhead{($^\circ$:$'$:$''$)}&
\colhead{($^\circ$)}&
\colhead{(min)}&
\colhead{FWHM}&
\colhead{Slits}&
\colhead{Target Velocities}&
\colhead{Velocities of}\\
\colhead{}&
\colhead{}&
\colhead{}&
\colhead{}&
\colhead{}&
\colhead{}&
\colhead{}&
\colhead{}&
\colhead{(Success Rate)}&
\colhead{Serendipitously Detected}\\
\colhead{}&
\colhead{}&
\colhead{}&
\colhead{}&
\colhead{}&
\colhead{}&
\colhead{}&
\colhead{}&
\colhead{}&
\colhead{Stars}
}
%Slit counts: without alignment or guide stars
%Star counts: including zq = 1, 3, 4
\startdata
	{M32\_1\tablenotemark{a}} &  2007 Nov 14 & 00 42 38.28 & $+$40 51 34.0  & $+$160.0
	& 2$\times$20 & $0\farcs5$ & 199 & 188 (94\%) & ~~72\\
	M32\_2 &  2008 Aug 03 & 00 43 03.82 & $+$40 55 07.7 & $+$70.0 
        & 3$\times$20 & $0\farcs6$ & 189 & 166 (88\%) & ~~27\\
	M32\_3 &  2008 Aug 03 & 00 43 11.60 & $+$40 52 34.7 & $-$110.0 
	& 3$\times$20 & $0\farcs7$  & 203 & ~132 (65\%)\tablenotemark{b}& ~~10\\
	M32\_4 & 2008 Aug 04 & 00 42 13.87 & $+$40 54 44.2 & $+$105.0
	& 3$\times$20  & $0\farcs6$ & 165 & 137 (83\%) & ~119\\
	M32\_5 &  2008 Aug 04 & 00 42 13.88 & $+$40 52 02.6 & $-$75.0 
	& 3$\times$20  & $0\farcs6$ & 177 & 152 (86\%) & ~~72 \\
        M32\_6 & 2008 Aug 31 & 00 41 20.41 & $+$41 51 32.2 & 0.0
        & 3$\times$20 $+$ 1$\times$10 & $0\farcs7$ & 169 & 152 (90\%) & ~128\\
        [1mm]

        SE7 & 2008 Sept 01 & 00 43 38.74 & $+$41 10 17.4 & $+$39.0
        & 2$\times$10 & $0\farcs8$ & 170 & 148 (87\%) & ~~50 \\
        SE8 & 2008 Sept 30 & 00 44 00.82 & $+$41 09 27.1 & $-$113.0
        & 3$\times$15 & $0\farcs5$ & 197 & 178 (90\%) & ~~27 \\
        SE9 & 2008 Oct 01 & 00 44 49.26 & $+$41 03 27.6 & $-$60.0
        & 2$\times$12.5 $+$ 1$\times$15 & $0\farcs4$ & 204 & 185
        (91\%) & ~~11 \\ [1mm]

        mctA5 & 2010 Oct 07 & 00 44 18.33 & $+$41 39 28.1 & $+$270.0 
        & 3$\times$16 & $0\farcs6$ & 212 & 197 (93\%) & ~~82 \\
        mctB4 & 2010 Oct 08 & 00 44 29.04 & $+$41 35 10.0 & $+$270.0
        & 3$\times$16 & $0\farcs7$ & 177 & 172 (97\%) & ~~81 \\ 
        mctC3 & 2010 Oct 07 & 00 45 11.89 & $+$41 53 37.7 & $+$90.0
        & 3$\times$16 & $0\farcs5$ & 198 & 170 (86\%) & ~~69 \\ 
        mctD3 & 2010 Oct 07 & 00 45 09.46 & $+$41 49 08.8 & $+$90.0
        & 3$\times$18 & $0\farcs6$ & 209 & 185 (84\%) & ~~69 \\
        mctE3 & 2010 Oct 08 & 00 46 53.23 & $+$42 14 59.3 & $+$90.0
        & 3$\times$17 & $0\farcs7$ & 221 & 202 (91\%) & ~~18 \\ [1mm]

        mct04p & 2010 Oct 07 & 00 44 51.81 & $+$41 25 19.2 & $-$142.3
        & 3$\times$16 & $0\farcs6$ & 254 & 223 (88\%) & ~~24 \\
        mct05p & 2010 Oct 08 & 00 44 19.70 & $+$41 32 53.5 & $-$52.3
        & 3$\times$16 & $0\farcs5$ & 254 & 188 (74\%) & ~~100 \\
        mct06p & 2010 Oct 07 & 00 44 33.71 & $+$41 36 17.6 & $-$52.3 
        & 3$\times$16 & $0\farcs5$ & 251 & 211 (84\%) & ~~43 \\
        mct07p & 2010 Oct 08 & 00 44 41.77 & $+$41 40 03.0 & $-$52.3
        & 3$\times$16 & $0\farcs6$ & 264 & 210 (80\%) & ~~71 \\
        %mct08p\tablenotemark{c} & 2010 Oct 07 & 00 44 54.88 & +41 43 04.8 & $-$52.3
        %& $3 \times 16$ & $0\farcs8$ & 258 & --- & --- \\ 
        mct09p & 2010 Oct 08 & 00 45 39.23 & $+$41 38 39.1 & $-$142.3 
        & 3$\times$16 & $0\farcs6$ & 252& 213 (85\%) & ~~22 \\
        mct10p & 2010 Oct 07 & 00 45 08.24 & $+$41 46 19.2 & $-$52.3 
        & 3$\times$18 & $0\farcs9$ & 255 & 207 (82\%) & ~~34\\
        mct12p & 2010 Oct 07 & 00 45 28.34 & $+$41 53 23.3 & $-$52.3 
        & 3$\times$18 & $0\farcs9$ & 265 & 212 (80\%) & ~~12\\
        mct13p  & 2010 Oct 08 & 00 45 42.02 & $+$41 56 42.4 & $-$52.3 
        & 3$\times$18 & $0\farcs7$ & 259 & 217 (84\%) & ~~23\\
        mct15p & 2010 Oct 08 & 00 45 54.36 & $+$41 59 43.1 & $-$52.3 
        & 3$\times$17 & $0\farcs9$ & 261 & 206 (79\%) &~~10\\
        mct16p & 2010 Oct 08 & 00 46 08.44 & $+$41 02 58.6 & $-$52.3
        & 3$\times$18 & $0\farcs7$ & 258 & 221 (86\%) & ~~~5\\ [1mm]
%for the mct slitmasks, "usable target velocities" have zquality ge 3. 
%'usable serendip velocities' include serendips from both the zspec
%structure and the serendip list; all have zquality ge 3. 
\tableline\tableline \vspace{1 mm}
        \\
	Total: &&&&&&& 5263~ & 4472 (85\%)~ & 1179
\enddata
%\tablecomments{Units of right ascension ($\alpha$) are in hours,
%  minutes and seconds.  Units of declination ($\delta$) are in
%  degrees, arcminutes and seconds.}
\tablenotetext{a}{The ``M32\_1'' slitmask was originally named ``M32'' at
  the time of submission of the slitmask design.}
\tablenotetext{b}{Due to a warp in the slitmask during
  the time of observation, approximately $30\%$ of the slitlets from
  mask M32\_3 did not produce useful spectra.}
%\tablenotetext{c}{The mct08p slitmask fails to go through reduction
 % process, so we do not include it in the analysis. The data quality
%  appears good to the eye. }
\label{tab_obs}
\end{deluxetable*}
\subsection{Source Catalogs}\label{ssec_catalogs}

All targets are chosen from an $i^{\prime}$-band $2^{\circ} \times
2^{\circ}$ CFHT/MegaCam mosaic centered on M31, obtained in
November 2004. We run the software package DAOPHOT \citep{ste94}
on the image to identify sources, fit PSFs, and produce a PSF-subtracted
residual image. The final catalog consists of nearly 2 million unique
sources. 

For our 2010 Keck/DEIMOS observing run, we also had access to data
from the first round of observations 
of the PHAT program. The data are organized into $12\farcm0 \times 6\farcm5$
bricks; three half-bricks were available at the time of our slitmask
design. From
these data, we created lists of different stellar populations: metal-poor ([Fe/H] $\lesssim -1.3$), metal-intermediate
($-1.3 \lesssim$ [Fe/H] $\lesssim -0.7$), and metal-rich ([Fe/H]
$\gtrsim -0.7$) RGB stars, and 
and hot, massive main sequence stars selected on the basis of SED
fitting to six-filter HST photometry. 
Though we do not treat stars differently based on subcategory
membership in the present paper, the kinematics of these stellar
populations and their relationship to the kinematics of the RGB
populations will be presented in a future work.

\subsection{Isolated Target Selection}\label{ssec_target}
Not all the sources in our catalogs are equally good candidates for
multi-object spectroscopy; we prefer isolated targets, those whose
spectra are least likely to be contaminated by light from neighboring
objects. We classify this contamination as either {\em crowding} or {\em
  blending}. We define a crowded source as one
that has at least one neighbor detected by DAOPHOT that is bright and
close enough to potentially interfere with the spectrum of the source.
In contrast, we define a blended source as one that is identified by DAOPHOT as
a single source but for which visual inspection of the
PSF-subtracted image indicates that more than one object may be
present. 

Furthermore, we only target stars in the apparent magnitude range
$20 < i^{\prime} < 22$ for M32 and SE slitmasks (modified to $20 < i^{\prime} < 21.5$ for MCT
slitmasks). We select a bright limit of $i^{\prime}=20$ because the tip of the red giant branch is at
$i^{\prime}=20.5$ in M31 and so the MW contamination fraction
increases significantly in brighter stars. The faint-end limit is
chosen because in very crowded
areas, it is difficult to recover high-quality spectra of stars
fainter than about $i^{\prime} = 22$. The surface density of sources
in the area covered by the MCT slitmasks is so high that we can efficiently pack targets on
our slitmasks even with a conservative faint-end limit of
$i^{\prime}=21.5$. 

To choose targets for all slitmasks, we first sort possible targets into
three lists (1, 2, 3) in decreasing order of isolation, with list 3
reserved for rejects (\S\,\ref{sss_secrowd} --
\S\,\ref{sss_mctblend}). Within each list, we prioritize
possible targets by magnitude, giving highest priority to
intermediate-brightness stars with $20.5 < i^{\prime} < 21.0$. In the
final target selection process, we exhaust each list before moving on
to the next. \S\,\ref{sss_hsttargets} describes additional selection
criteria specific to the five PHAT-based slitmasks. 

\subsubsection{Crowding in the SE \& M32 Data Sets}\label{sss_secrowd}
In the M32 and SE data sets, we use a neighbor-rejection test to
eliminate crowded sources. We reject any star with at least one
neighbor with a sufficient combination of proximity and relative
brightness, i.e., satisfies the following empirical criterion
determined from visual inspection of the CFHT/MegaCam image: 

\begin{equation}
I_{\rm nbr} < I_{\rm tgt} - \Bigg(\frac{d}{0\farcs8}\Bigg)^2 + 3.0
\label{eq1}
\end{equation}

\noindent Here, $I_{\rm tgt}$ and $I_{\rm nbr}$ are the $i^{\prime}$-band
magnitude of the target source and the neighbor, respectively, and $d$ is the
distance in arcseconds between the target and the neighbor. This cut
eliminates about $90\%$ of the stars in the M32 and SE data sets \citep{how12}.

\subsubsection{Blending in the SE \& M32 Data Sets}\label{sss_seblend}
We identify likely blends in the SE and M32 data sets by visually
inspecting the high-pass filtered and PSF-subtracted versions of the
$i^{\prime}$-band CFHT/MegaCam image at the locations of the stars that
survive the crowding test. Each target is flagged as unblended,
marginally blended, or badly blended depending on the degree to which
its image resembles the PSF on the high-pass filtered image and the
strength of systematic residuals at its location on the residual image.

\subsubsection{Crowding in the MCT Data Set}\label{sss_mctcrowd}
Based on our experience with the M32 and SE spectroscopic data sets,
we decide to use a slightly relaxed crowding criterion for the MCT
data set, rejecting (assigning to list 3) any catalog entry with at
least one neighbor which satisfies the following: 

\begin{equation}
I_{\rm nbr} < I_{\rm tgt} - \Bigg(\frac{d}{0\farcs8}\Bigg)^{3/2}+ 3.0
\end{equation}

\noindent This change accounts for the fact that the seeing during our
spectroscopic observations tends to be slightly better than the
$0\farcs8$ CFHT seeing upon which the original criterion was based. This
cut eliminates $80\%-90\%$ of the stars in the inner slitmasks, $50\%-70\%$
at intermediate radii, and only $40\%$ in the furthest slitmask, mctE3. 

\subsubsection{Blending in the MCT Data Set}\label{sss_mctblend}
In order to avoid a tedious visual inspection of the large area covered by the MCT
data set, we design two empirically-based statistical tests to
detect possible blends. We visually inspect and flag as ``blended'' or
``non-blended'' $\lesssim100$ objects in each of three small
representative image sections at 
different distances from the center of M31. We then design
quantitative tests that approximately reproduce our visual
classifications. 

\begin{figure}[h]
\scalebox{1}{\includegraphics[trim=70 375 65 95, clip = true, width=0.48\textwidth]{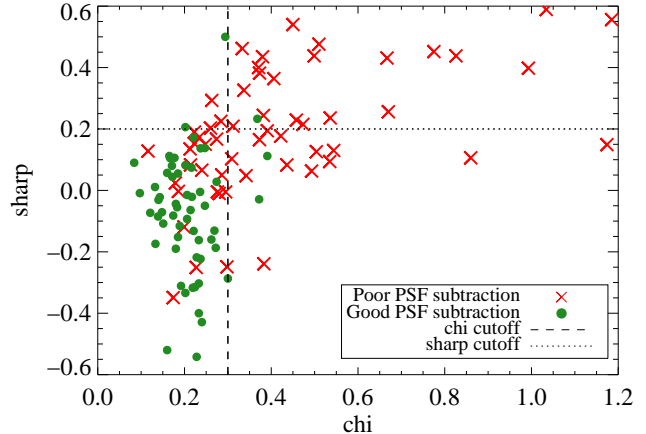}}
\centering
\caption{{\tt chi/sharp} isolation criteria. Visually identified
  blends (red crosses) and non-blends (green circles) are shown from a
  small representative area close to the center of M31. We reject
  all sources with DAOPHOT goodness-of-fit parameters {\tt chi}
  \textgreater~0.3 (to the right of the  dashed line) or shape
  parameter {\tt sharp} \textgreater~0.2 (above  the dotted line). The
  {\tt chi} cutoff value is lowered in less  dense fields farther from
  the galactic center.}
\label{fig_chisharp}
\end{figure}

The first test is based on the DAOPHOT-generated goodness-of-fit
parameter {\tt chi} and shape parameter {\tt sharp}.  Objects that appear isolated based
on visual inspection fall into a well-defined locus in {\tt chi/sharp}
space, as shown in Figure~\ref{fig_chisharp}. Based on this
relationship, we retain only objects with {\tt
  sharp} $<0.2$. We assign a radially dependent linear cut in {\tt
  chi}, accepting all stars with {\tt chi} $ <0.3$ in the crowded
areas and progressing to the more stringent criterion {\tt chi}
$<0.2$ in the least crowded outermost slitmask. This cut eliminates $20\%$ of the
remaining candidate targets. Because the dividing line between
isolated sources and likely blends is less well defined in {\tt chi}
than {\tt sharp}, we allow a buffer zone $0.5$ units wide in {\tt
  chi}. The $1\%$ of stars in the buffer zone are relegated to
list 2, but not rejected outright. 

\begin{figure}[h]
\scalebox{1}{\includegraphics[trim=70 370 65 95, clip = true,
  width=0.48\textwidth]{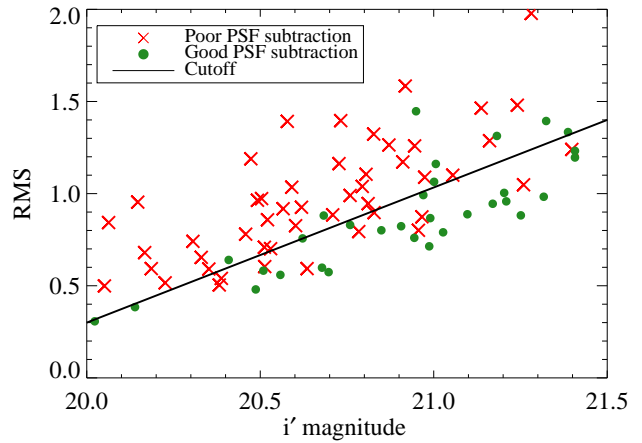}}
\centering
\caption{RMS isolation criterion. Visually identified blends (red
  crosses) and non-blends  (green circles) are shown from a small
  representative area about 8~kpc from the center of M31. The RMS
  deviation of the pixel values of the residual image in a $5\times
  5$ pixel square centered on each source is plotted against the
  $i^{\prime}$ magnitude of the source. We reject all sources above
  the solid black line.}
\label{fig_rms}
\end{figure}

In the second test for possible blends, we compare the apparent
quality of subtraction to the normalized RMS flux of a $5
\times 5$ pixel square of the PSF-subtracted image centered on the
source. This value tends to increase with apparent degree of blending. We
determine that the best cut is a linear function of magnitude,
where blended sources have $\frac{\rm RMS}{<\rm flux>} > 0.3$ at
$i^{\prime}=20$ and $\frac{\rm RMS}{<\rm flux>} > 1.2$ at
$i^{\prime}=21.5$ (Figure~\ref{fig_rms}). The
36\% of stars that fail this test are flagged as ``possibly
isolated'' and pushed to list~2. Because the
correlation between apparent PSF subtraction quality and RMS is not as
tight as those in the {\tt chi/sharp} test, we do not use the RMS cut
to reject (assign to list~3) targets that are ``isolated" according to both the
neighbor-rejection and {\tt chi/sharp} tests. 

\subsubsection{Target Selection for PHAT-based Masks}\label{sss_hsttargets}
We also design five slitmasks based jointly on CFHT data and the
PHAT survey-based stellar population lists described in
\S\,\ref{ssec_catalogs}. We require that a star from the PHAT catalog be
in the CFHT catalog and have passed through the filters described
above to be considered for selection. To ensure final selection of the
most isolated (list 1) PHAT
objects, we push all non-PHAT objects down one list. For these five
slitmasks, then, the priority scheme is as follows:  list~1 consists
of only isolated, PHAT-selected sources; list~2 includes the isolated
CFHT-only sources; and list~3 includes all of the possibly-isolated
sources. Because the shape of the PHAT survey  bricks is different
from the shape of DEIMOS slitmasks, large areas of the DEIMOS
slitmasks that overlap PHAT survey bricks do not target PHAT-based
sources. In these areas, the mask design software automatically
proceeds to list~2 to select CFHT-based objects.

\subsubsection{Summary of Target Selection}
To summarize, each source passes through three tests for isolation:
neighbor-rejection, {\tt chi/sharp}, and RMS. Each source receives a
score for each test: 0 if isolated, 0.6 if marginally isolated, and 2.0
if not isolated. The sum of the three scores determines which list the
source belongs in. A total score of 0 maps to list 1; a total of less
than 2.0 maps to list 2; and a star with a total score of 2.0 or
greater is assigned to list 3. Hence, any star that
fails either the neighbor-rejection or {\tt chi/sharp} test cannot be
selected, and any star that passes all three filters is given highest
priority. 

None of the methods described here can fully eliminate the possiblity of
placing one slit over several objects. Especially in the crowded inner
areas, it is very common to obtain mutiple spectra in
one slit. We discuss our handling of these serendipitous detections in
\S\,\ref{ssec_serendip}. In addition, a small percentage of the target spectra may
still be contaminated with light from nearby objects; objects with
unusable spectra are identified by eye and removed from the sample at
the end of the data reduction process as described in
\S\,\ref{ssec_quality}. The target selection process outlined above
serves simply to make educated guesses about the objects best suited
for spectroscopy. 

In \S\,\ref{ssec_kstest} we show that the spectroscopic target selection criteria
(PHAT CMD vs. magnitude only) and actual degree of crowding (isolated
vs. sharing a slit with another bright object) have minimal, if any,
effect on the measured velocity distribution. 

%Figure 2

%Figure 3

\vspace{4mm}
\subsection{Observations}\label{ssec_multi}
 
All slitmasks are observed using Keck/DEIMOS with the 1200 line mm$^{-1}$
grating. This configuration yields a spatial scale of $0\farcs12$
 pixel$^{-1}$ and a spectral dispersion of $0.33$ \AA\ pixel$^{-1}$. 
 We set the central wavelength to $7800$ \AA,  corresponding to a
 wavelength range of $\sim 6450-9150$ \AA.  The exact 
wavelength range for each slit varies as a result of location on
 the slitmask and/or truncation due to vignetting.  The 
wavelength region is chosen to target the Ca {\scriptsize\rm{II}} 
triplet absorption feature present in RGB stars.  The anamorphic
 distortion factor for this grating and central wavelength is 0.606.
 Therefore, each $0\farcs8$ wide slitlet subtends $4.1$ pixels. 
 Better still, excellent seeing conditions ($\sim0\farcs6$) 
during observations can provide somewhat better spectral resolution
 yielding an average resolution of $3.1$ pixels $= 1.0$ \AA. 

Reliable spectra (those that yield secure velocities, as described in
\S\,\ref{ssec_quality}), are obtained from 4465 of the
 5263 slitlets. Approximately $30\%$ of the slitlets from slitmask M32\_3 
did not produce useful spectra due to a warp in the slitmask. 

See Table~\ref{tab_obs}  for information on the positions of the slitmasks and
the number of useful spectra recovered from each one. 

\subsection{Data Reduction}\label{ssec_mreduce}
% Spec2d automatically extracts spectra, spec1d does cross correlation
% and \textit{zspec} can re-extract and refit 
 
The Keck/DEIMOS multiobject slitmasks are processed using the 
\textit{spec2d} and \textit{spec1d} software (version 1.1.4) developed 
by the DEEP Galaxy Redshift Survey team at the University of
California,
Berkeley \citep{dav03}\footnote{http://astro.berkeley.edu/$^{\sim}$cooper/deep/spec2d/}.  
Briefly, the reduction pipeline rectifies, flat-field and fringe
corrects, wavelength calibrates, sky subtracts, and cosmic ray 
cleans the two-dimensional spectra, and extracts the one-dimensional 
spectra. For more details, see \citet{how12}.

 \subsection{Cross-Correlation Analysis}\label{ssec_cc}

Line-of-sight (LOS) velocities for resolved targets are measured from the
one-dimensional spectra using a \citet{geh10} modified version 
of the visual inspection software \textit{zspec}, developed by D.
 Madgwick  for the DEEP Galaxy Redshift Survey at the University 
of California, Berkeley. The software determines the best-fit LOS 
velocity for a target by cross-correlating its one-dimensional 
science spectrum with high signal-to-noise stellar templates
 in pixel space and locating the best fit in reduced-$\chi^2$ space.  

A-band telluric corrections and heliocentric corrections are
calculated and applied to the measured LOS velocities.  The
 A-band telluric corrections, which account for velocity errors 
associated with the slight mis-centering of a star in a slit, are
determined using the method discussed in \citet{soh07} and
\citet{sim07}. 

LOS velocity errors are determined for each star by adding in quadrature
the cross-correlation based velocity error and a systematic error
estimated by repeat observations, as described in \citet{how12}. The
typical LOS velocity error in our sample is 4--5~km~s$^{-1}$.  

%---------------------------------
% 2.7. QUALITY ASSESSMENT
%---------------------------------
\subsection{Quality Assessment}\label{ssec_quality}

Each two-dimensional spectrum, one-dimensional spectrum, and 
corresponding Doppler shifted template match are visually inspected 
in \textit{zspec} and assigned a quality code based on the 
reliability of the fit.  This process allows the user to evaluate the
quality of a spectrum and reject instrumental failures and poor
quality spectra.   Velocity measurements based on two or more 
strong spectral features are labeled ``secure.'' Velocity
measurements based on one strong feature plus additional marginal
features are labeled ``marginal.''  Spectra that contain no strong
features, low S/N and/or instrumental failures are considered
unreliable, and so are not included in our analysis.  Additional
details on quality code assignment can be found in
\citet{guh06}. During this process, we also identify and flag $43$ likely
MW M dwarfs based on their strong surface-gravity sensitive Na I
8190A doublet. These are excluded from the radial
velocity analysis.    

%---------------------------------
% 2.8. SERENDIPITOUS SOURCES
%---------------------------------
\vspace{3mm}
\subsection{Serendipitous Sources}\label{ssec_serendip}

Upon visual inspection of the one-dimensional and two-dimensional
 spectra during the quality assessment phase outlined in
 \S\,\ref{ssec_quality}, some fraction of the slits clearly show that 
the slitlet intersects more than one star: the
target star and one or more serendipitously detected stars, or {\em
  serendips}. Serendips are detected via one of two methods: through
continuum detections that are offset from the primary target in the
spatial direction, or by the detection of spectral features that are
offset from the primary target in the spectral direction. Serendip 
detections occur frequently in the inner parts of M31 and close to M32 due
to the severe crowding  and blending in the CFHT/MegaCam data.
Serendips are also assigned quality codes and those with secure or marginal
velocities are included in the radial velocity analysis. More
information on  serendipitous detections can be  found in
\citet{how12}. 

 %-----------------------------------------------------------
 %  3. DATA ANALYSIS
 %-----------------------------------------------------------
\section{DATA ANALYSIS}\label{sec_analysis}

We perform our analysis in each of the five spatial regions labeled in
Figure~\ref{fig_allsubregions}. These regions are the SE minor
axis (not expected to yield constraints on the  spheroid rotation velocity); the SSW
quadrant (expected to yield a constraint on the rotation velocity via
a negative velocity offset from the systemic velocity of M31); and three
regions along the northeast major axis, which we name NE1, NE2 and
NE3 in order of increasing projected radial distance (expected to yield three independent
estimates of the rotation velocity via positive velocity offsets from the
systemic velocity of M31). Note that these regions are
defined by lines of constant position angle and projected radius, and
are not quite the same as the three data sets shown in
Figure~\ref{fig_galex}. The positions of and number of stars within
each region are given in the first six columns of Table~\ref{tab_results}.

The kinematically cold peak at around $-100~\rm km~s^{-1}$ in Figure~\ref{fig_mcthist} suggests that
our measured velocity distribution has a significant contribution from
the disk. Hence, it is imperative that we realistically account for
disk contributions. Instead of adopting a specific model for its
velocity field, we only assume that the stellar disk is locally cold with a symmetric
velocity distribution.  As explained in \S\,\ref{ssec_bins}, we apply
this assumption to divide each region into several
subregions. To each subregion we fit two Gaussian distributions,
corresponding to a kinematically cold and a hot component, where the
hot component is required to have the same mean velocity and velocity
dispersion across all subregions in a region (described in
\S\,\ref{ssec_fits}). Lastly, in \S\,\ref{ssec_gss} we
describe how we modify our analysis to account for the possibility 
of contamination by the GSS and associated tidal debris.

\subsection{Choice of Subregions and Expected Disk LOS Velocity
  Pattern}\label{ssec_bins}

Because we assume that the disk is only locally cold, we fit
for the disk in each of many small subregions. The spatial boundaries
of individual subregions are dictated by two competing desirable
factors, namely a small spread in disk mean velocity and high number statistics. Our
assumption of a perfectly cold disk is only strictly true in the limit
of infinitely small subregions. On the other hand, a multi-Gaussian
fit to a velocity distribution requires a somewhat large number of
points. We arbitrarily decide that $100$ is the minimum satisfactory
number of points.

To estimate the spread of the mean disk velocity as a function of
position, we employ a simple geometrical model for the rotation pattern of an
inclined disk with perfectly circular motion \citep{guh88}:

\begin{equation}
v_{\rm obs}(\xi, \eta) = v_{\rm sys} \pm  \frac{v_{\rm rot}(r_{\rm
    deproj}) \sin(i)}{ \sqrt{1 +
    \tan^2(\Delta {\rm PA}) / \cos^2(i) }}
\label{eq_rot}
\end{equation}

\noindent where $\xi,\eta$ are tangent-plane coordinates with origin
at the center of M31, $i=77^{\circ}$ is the inclination of the disk of M31, $v_{\rm
  sys} = -300 \rm~km~s^{-1}$ is the systemic heliocentric velocity of
M31, $\Delta \rm PA$ is the position angle projected  onto
the plane of the sky measured relative to the major axis of the disk
of M31, $r_{\rm deproj}$ is the radial position measured in the plane
of the disk, $v_{\rm rot}$ is the disk rotation speed, and the $+$ and $-$ signs apply to the NE and  SW halves
of the disk, respectively. An azimuthally averaged estimate for $v_{\rm  rot}(r_{\rm deproj})$, based on HI
kinematics, ranges from $\sim 250~\rm km~s^{-1}$  at $r_{\rm   deproj}
= 15~ \rm kpc$ to $\sim 175\rm ~km~s^{-1}$ at  $5\rm~kpc$
\citep{cor10}. The expected velocity spread calculated in this way
will not be exact, because in addition to $\Delta \rm PA$, 
the true spread in mean stellar velocity over a subregion is
influenced by $\Delta r_{\rm deproj}$ of the subregion, any
departure from perfectly circular motion, and the intrinsic local
velocity distribution (due to a multiple-component disk, for
example).
 
\begin{figure}[h]
\scalebox{1}{\includegraphics[trim=10 50 50 160, clip = true, width = 0.48\textwidth]{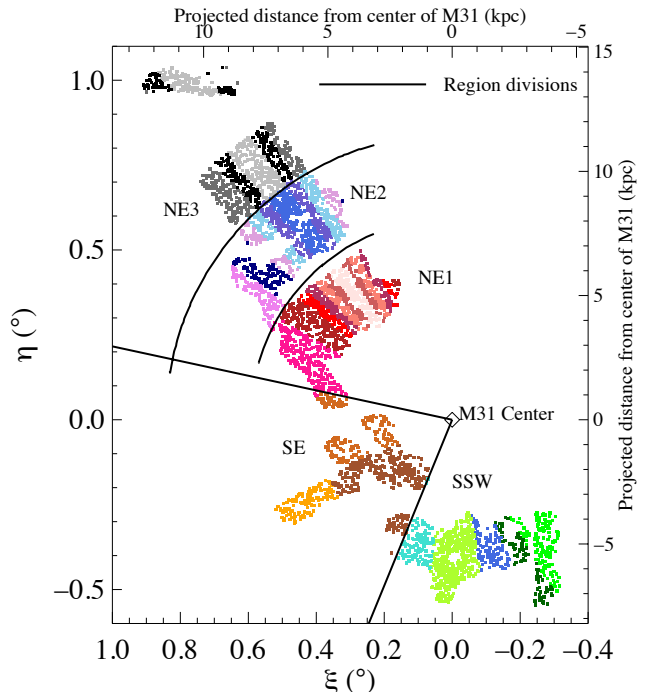}}
\centering
\caption{Division into regions and subregions. Each point corresponds
  to a single  velocity measurement. Solid lines delineate the five
  regions  for which we obtain independent kinematical parameters of
  the spheroid: inner northeastern (NE1), intermediate northeastern
  (NE2), outer northeastern (NE3), southeastern (SE), and
  south-southwestern (SSW). Colors show the subregions within each
 region that are used to determine the disk contribution to the
  velocity distribution. With the exception of the SE region, our
  subregion naming convention is based on distance from the major
  axis. For example, in the NE3 region, $\rm NE3_1$ (light gray) straddles
  the major axis, $\rm NE3_2$ (black) is slightly farther out, and $\rm NE3_3$
  (dark gray) is farthest from the major axis. In the
  SE region, we use projected distance from the center of M31 in
  addition to lines of constant position angle to define three
  subregions: $\rm SE_1$ (outer, yellow), $\rm SE1_2$ (inner south, dark brown)
  and $\rm SE1_3$ (inner north, light brown).} 
\label{fig_allsubregions}
\end{figure}

\begin{figure}[h]

\scalebox{1}{\includegraphics[trim=110 30 35 475, clip = true, width = 0.48\textwidth]{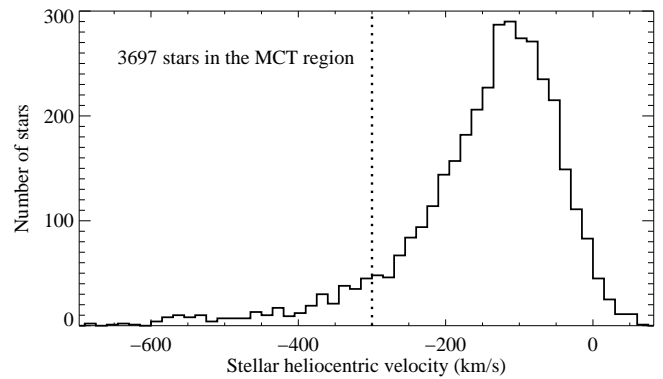}}
\centering
\caption{Distribution of heliocentric radial velocities for all 3697
  stars with reliable velocity measurements from the MCT
  slitmasks. The peak around $-$100$~\rm km~s^{-1}$ corresponds to the
  disk, the LOS projection of whose rotation speed causes stars to be
  offset by as much as $+$200$~\rm km~s^{-1}$ with respect to
  the $-$300$~\rm km~s^{-1}$ systemic velocity of M31 (dotted line).}
\label{fig_mcthist}

\end{figure}

Using Equation~\ref{eq_rot} with $v_{\rm rot} = 250  \rm~km~s^{-1}$,
we bin our data in each region into subregions based on position
angle. The angle subtended by a single subregion is approximately the
greater of two $\Delta \rm PA$ criteria: 1) the $\Delta \rm PA$ such that the change in $v_{\rm obs}$
over a subregion due to $\Delta \rm PA$ is $10 \rm ~km~s^{-1}$, or 2)
the $\Delta \rm PA$ that includes 100 data points. Our final subregions are
shown in Figure~\ref{fig_allsubregions}.  In the NE1--NE3 and SSW regions,
we identify these subregions
with subscripts that increase with distance from the nearest major
axis: $\rm NE1_1, NE1_2, \ldots, NE1_7;~SSW_1, \ldots, SSW_5, etc.$ In the SE
region, the outer, inner south, and inner north subregions are named
$\rm SE_1, SE_2, SE_3,$ respectively. Note that we use this rotation
pattern only to estimate appropriate bin sizes, not to determine disk
rotation speeds. The final positions of and number of stars in each
subregion are presented in the first six columns of
Table~\ref{tab_disk} in the Appendix.

%FIGURE 4

%Figure 5
\subsection{Fitting the Velocity Distribution Model}\label{ssec_fits}

\newcommand{\cold}{\text{disk}}
\newcommand{\hot}{\text{sph}}
\newcommand{\barv}{\overline{v}}
\newcommand{\gaussian}[3]{\mathcal{N}\!\left(#1 \,\big\vert\, #2, \, #3\right)}

\newcommand{\rj}{r(j)}
\newcommand{\sj}{s(j)}
\newcommand{\kmpersec}{\rm ~km~s^{-1}}
\newcommand{\eqnref}[1]{Equation~\ref{#1}}
\newcommand{\code}[1]{\emph{#1}}

We use a Markov chain Monte Carlo (MCMC) sampler to find the velocity
distributions of the disk and spheroid in each region.  The spheroid
is modeled as a single Gaussian distribution for each region, while
the disk is modeled as a Gaussian distribution for each subregion.  
The model likelihood at each point in parameter space is then:
\begin{eqnarray}\displaystyle
  L = \prod_{j=1}^{N}
  & \Big[
    &
    f_{\sj}
    \, \gaussian{v_{j}}{\barv_{\hot,\rj}}{\sigma_{\hot,\rj}} +
    \nonumber \\
    &&
    \left( 1 - f_{\rj} \right) \gaussian{v_{j}}{\barv_{\cold,\sj}}{\sigma_{\cold,\sj}}
    \Big]
    \label{eqn:likelihood}
\end{eqnarray}
where $N$ is the number of stars in our data set, and star $j$ with
measured velocity $v_j$ is
found in region $\rj$ and subregion $\sj$.  The notation
$\gaussian{x}{\mu}{\sigma}$ indicates the Gaussian distribution with
mean $\mu$ and standard deviation $\sigma$ evaluated at $x$.  The
scalar $f_{s}$, subject to $0 \le f_{s} \le 1$, is the fraction of the
stars in subregion $s$ that belong to the spheroid.  The spheroid in
region $r$ has mean velocity $\barv_{\hot,r}$ and velocity dispersion
$\sigma_{\hot,r}$, while the disk in subregion $s$ is characterized by
velocity $\barv_{\cold,s}$ and dispersion $\sigma_{\cold,s}$.

Subregion $\textrm{SSW}_4$ includes stars from the galaxy M32, so for
that subregion only we replace the single-Gaussian disk model
$\gaussian{v_{j}}{\barv_{\cold,\sj}}{\sigma_{\cold,\sj}}$ with a
double-Gaussian model.

The likelihood in \eqnref{eqn:likelihood}\ can be sampled
independently in each of the five regions.
We use the code {\tt emcee} \citep{for12},
which implements the affine-invariant ensemble sampler of
\citet{gw10}, to perform the MCMC algorithm.
In addition to the likelihood above, we must specify prior probability
distributions for the parameters.  For $f_s$ we use a uniform
distribution on the interval $[0, 1]$; for the mean velocities
$\barv_{\hot,r}$ and $\barv_{\cold,s}$ we use flat priors; and for the
dispersions $\sigma_{\hot,r}$ and $\sigma_{\cold,s}$ we demand
positive values.  
If we sample each region independently, then the parameter space for
region $r$ includes $\barv_{\hot,r}$, $\sigma_{\hot,r}$, plus the set
of $f_s$, $\barv_{\cold,s}$, $\sigma_{\cold,s}$ for each subregion $s$
within $r$.
We initialize the MCMC at $\barv_{\hot,r} = -300 \kmpersec$,
$\sigma_{\hot,r} = 150 \kmpersec$, each $f_s = \frac{1}{2}$, and
$\barv_{\cold,s}$ and $\sigma_{\cold,s}$  = the sample mean and
standard deviation of the velocities in subregion $s$.

Briefly, the MCMC ensemble sampler works as follows. It explores the
parameter space by maintaining a set of \emph{walkers}.  Each walker
represents a point in the parameter space.
At each iteration of the MCMC, each walker takes a step in the
parameter space by choosing another walker and stepping along the line
in parameter space connecting itself to the other walker.  The step
size is chosen stochastically and allows interpolation as well as
extrapolaton.  In effect, the walkers choose their steps based on the
covariance of the set of walkers.
After each step is taken, the posterior probability distribution at
the new point in parameter space is evaluated.  Steps that increase
the probability are always accepted, while steps that decrease the
probability are \emph{sometimes} accepted.
After a large number of steps, the ensemble of walkers will sample the
parameter space with frequency proportional to the posterior
probability distribution; we can draw fair samples from the
distribution by selecting points from the histories of the walkers.
We estimate the mean and variance of each parameter
(assuming unimodal distributions) based on sample statistics of the
histories of the walkers. In particular, we allow each of 32 walkers
to take 10,000 steps. We then compute the mean spheroid velocity as the mean
value of $\barv_{\hot,r}$ over the last 2000 steps of all the walkers
(i.e., the mean of 64,000 points), and the 68\%
confidence interval as the standard deviation of that quantity over
the last 2000 steps of all of the walkers. We use the same method to estimate the values and
uncertainties of the spheroid dispersion and the disk parameters. We
report these values in columns 7 and 8 of Table~\ref{tab_results} and
illustrate them in Figure~\ref{fig_profiles_nogss}. 

\subsection{Accounting for Tidal Debris Associated with the Giant
  Southern Stream}\label{ssec_gss}

The analysis described thus far separates the spheroid from the disk
(and from M32 in subregion $\text{SSW}_4$). The mean velocity and
dispersion we extract describe the average properties of the spheroid,
regardless of its underlying structure.

Several lines of evidence, however,
suggest that our measurements do not well represent a rotation curve
of a smooth spheroid. First, the mean spheroid velocities
in the five regions do not follow a physical rotation pattern: the
mean velocity of the NE2 region is consistent with zero, while
the surrounding regions (NE1 and NE3) are rotating at about $50~\rm
km~s^{-1}$ in the same
direction as the disk. Second, the sum of hot and cold Gaussians does not
well represent the velocity distributions. A chi-squared analysis
applied to the data binned by $20~\rm km~s^{-1}$ reveals that that the
probability of the model representing the velocity distribution is
relatively low (see the final column of
Table~\ref{tab_results}).

\begin{figure*}[h]
\begin{center}
\scalebox{1}{\includegraphics[trim=20 80 30 50,
  clip=true, width = 1\textwidth]{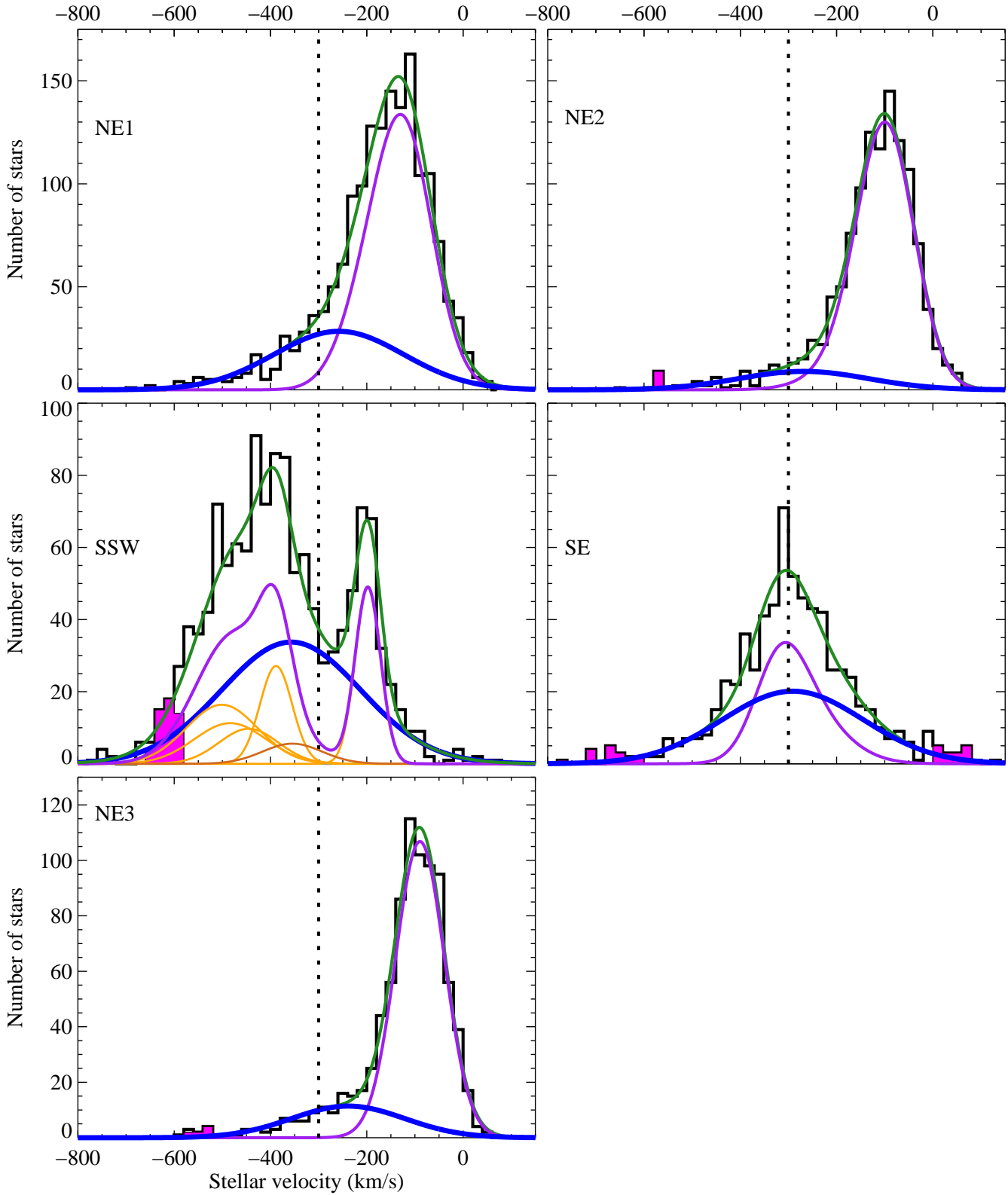}}
\centering
\caption{Maximum likelihood fits of a kinematically hot spheroid to
  each of the five regions in our sample, after excluding the velocity
  range encompassing the GSS and its
  associated tidal debris (shaded pink). Violet lines show the
  cumulative region cold component; blue show the best-fit spheroid
  Gaussian; green show the
  sum of these two components. The dashed lines show the systemic
  velocity of M31 relative to the MW. Individual subregion cold
  components are shown in orange in the SSW region panel, but left out
  of the other panels for clarity. We remind the
  reader that our fitting procedure makes use of the distinct velocity
profiles of the individual subregions as shown in
Figures~\ref{fig_NE1_subs} and 13-16, so that the fits are much better constrained
than may be apparent from the combined distributions here.}
\label{fig_6plots}
\end{center}
\end{figure*}

Velocity histograms, such as those in Figure~\ref{fig_6plots}, suggest that cold substructure, possibly
tidal debris from the GSS, could be skewing our
measurements. A close-up view of the negative-velocity tail of the NE2
histogram reveals a cold spike of about 10 stars in excess of the
Gaussian tail at $-580 \rm~ km~s^{-1}$ (Figure~\ref{fig_6plots}, top
right). A slight overdensity of points at this velocity can be seen in
the NE1 and NE3 histograms as well. Though nothing is immediately visible in the
SE or SSW regions, a few extra stars around $-580~\rm km~s^{-1}$ would
be partially concealed by the bulk of the velocity distribution.  

\begin{figure}[h]
\begin{center}
\scalebox{1}{\includegraphics[trim= 80 370 20 28, clip=true, width = 0.48\textwidth]{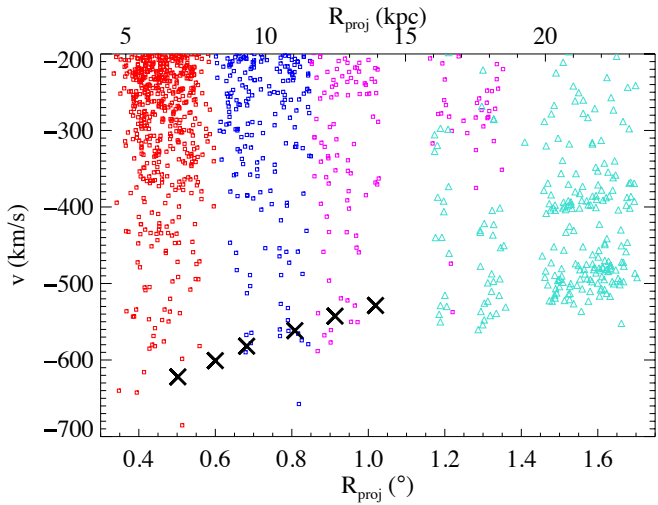}}
\centering
\caption{Radial velocity vs. projected radius of the most
  negative-velocity stars in the  NE1, NE2, and
NE3 regions (red, blue, and magenta, respectively). Overplotted are
  data from two fields centered on the GSS south of M31 (turquoise
 triangles) \citep{gil09}. The clusters of these turquoise
 triangles around $v \sim -500~\rm km~s^{-1}$ and $v \sim -390~\rm
 km~s^{-1}$ are the GSS and the secondary stream, respectively. The
 black crosses show six points for the predicted velocity of the NE
 Shelf \protect\citep{far12}. The GSS appears
to be present in the NE fields as slight concentrations around the
black crosses. }

\label{fig_v_rproj_all}
\end{center}
\end{figure}

Figure~\ref{fig_v_rproj_all} shows another projection
of these data: the velocity of each star in the NE1, NE2, and NE3
regions plotted against $R_{\rm proj}$. Also shown (in turquoise
triangles) are data from two GSS fields at 17 and
21 kpc \citep{gil09}. The concentrations of stars at approximately
$-500$ and $-400~\rm km~s^{-1}$ in these data represent the GSS and
the secondary stream, respectively. The magnitudes of the central stream velocities increase with decreasing $R_{\rm proj}$ as the
streams fall into the potential well of M31 from the south. The black
crosses in Figure~\ref{fig_v_rproj_all} show the expected stream
velocity as a function of radius closer to the center of M31
\citep{far12}. The NE region data from this work show clear
concentrations of objects near the predicted stream velocity,
continuing the trend seen south of the galaxy. Hence, it seems
unlikely that the peak in the Figure~\ref{fig_6plots} histograms is
simply a binning artifact, and probable that it comes from the
northern extension of a cold tidal stream. 

To account for the presence of the GSS and its associated tidal debris, we repeat the
MCMC fits in regions NE1, NE2, NE3, and SSW after removing all stars
within $\sigma = \pm 30 \rm~km~s^{-1}$ of the predicted velocities of
the two streams. (The measured velocity dispersion of the stream from
\citet{gil09} is $20~\rm km~s^{-1}$; however, we use the larger value
to ensure that we exclude all stream stars despite the slightly
uncertain mean stream velocity.) In the SE minor axis region, we
exclude all stars with velocity $v < -600~\rm km~s^{-1}$ or $v > 0~\rm
km~s^{-1}$. The former is to account for GSS debris, and the latter
for SE Shelf stars (GSS debris from third pericentric passage) \citep{gil07}. 

We account for the fact that we have removed stars within a range of
velocities by renormalizing the Gaussian distributions.  That is,
given a Gaussian velocity distribution, we compute the fraction of the
probability mass that falls within the excised velocity range and
scale up the remainder of the distribution so that the integral over
the remaining (unexcised) velocities is unity. The resulting
kinematical spheroid parameters are reported in columns 7 and 8 of
Table~\ref{tab_results_gss}. 

\begin{figure*}[h]
\scalebox{1.0}{\includegraphics[trim = 15 15 55 360, clip =
  true, width = 1\textwidth]{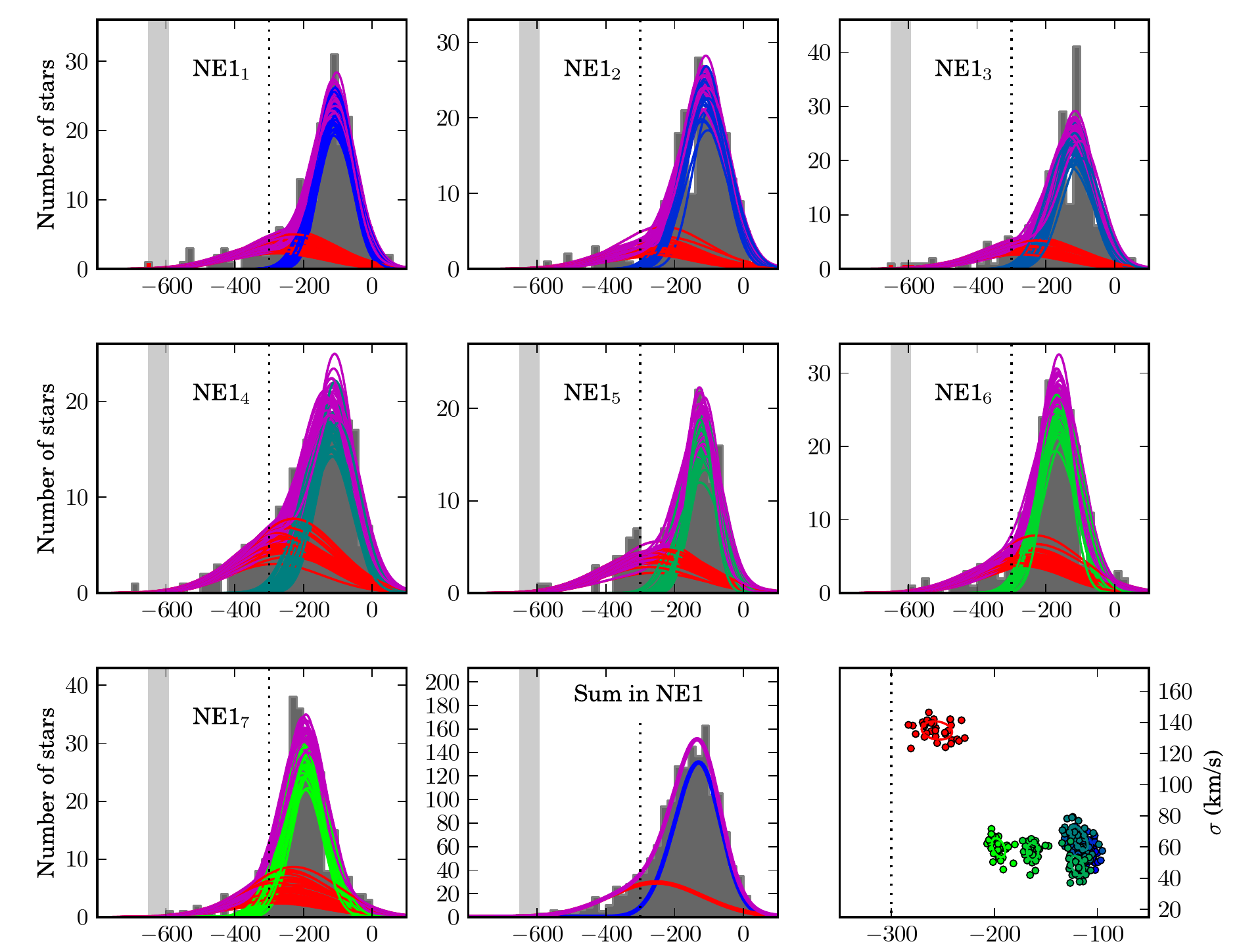}}
\centering
%\vspace{1 mm}
\caption{MCMC fits of kinematically hot (red) and cold (blue to green) components
  to each of 7 subregions in
  the NE1 (inner northeast major axis) region plotted over velocity histograms
  of stars in each subregion. Subregions progress from $\rm NE1_1,$
  straddling the
  NE major axis, to $\rm NE1_7,$ farthest from the major axis. One curve is
  drawn for each component for each of the 32 walkers in the
  MCMC. Each violet distribution is
the sum of the hot and cold distributions corresponding to a single walker. Velocity ranges excluded
due to possible contamination by tidal debris from the GSS are shown in two ways: by the gray shaded regions,
and by the stars shaded red on the histogram. The bottom middle panel
shows the best-fit hot component (red) that, when added to the
cumulative region cold distribution (blue), best fits the observed
velocity distribution. The bottom right panel shows the final position
of the walkers in parameter space. Ellipses show the mean and
uncertainty of each of the parameters $(\overline{v}, \sigma_v)$ for 
each of the kinematical subcomponents. The dotted line in each panel
marks the systemic velocity of M31 relative to the MW.}
\label{fig_NE1_subs}
\end{figure*}

\begin{figure*}[h]
\scalebox{1}{\includegraphics[trim=20 50 10 430, clip = true, width = 1\textwidth]{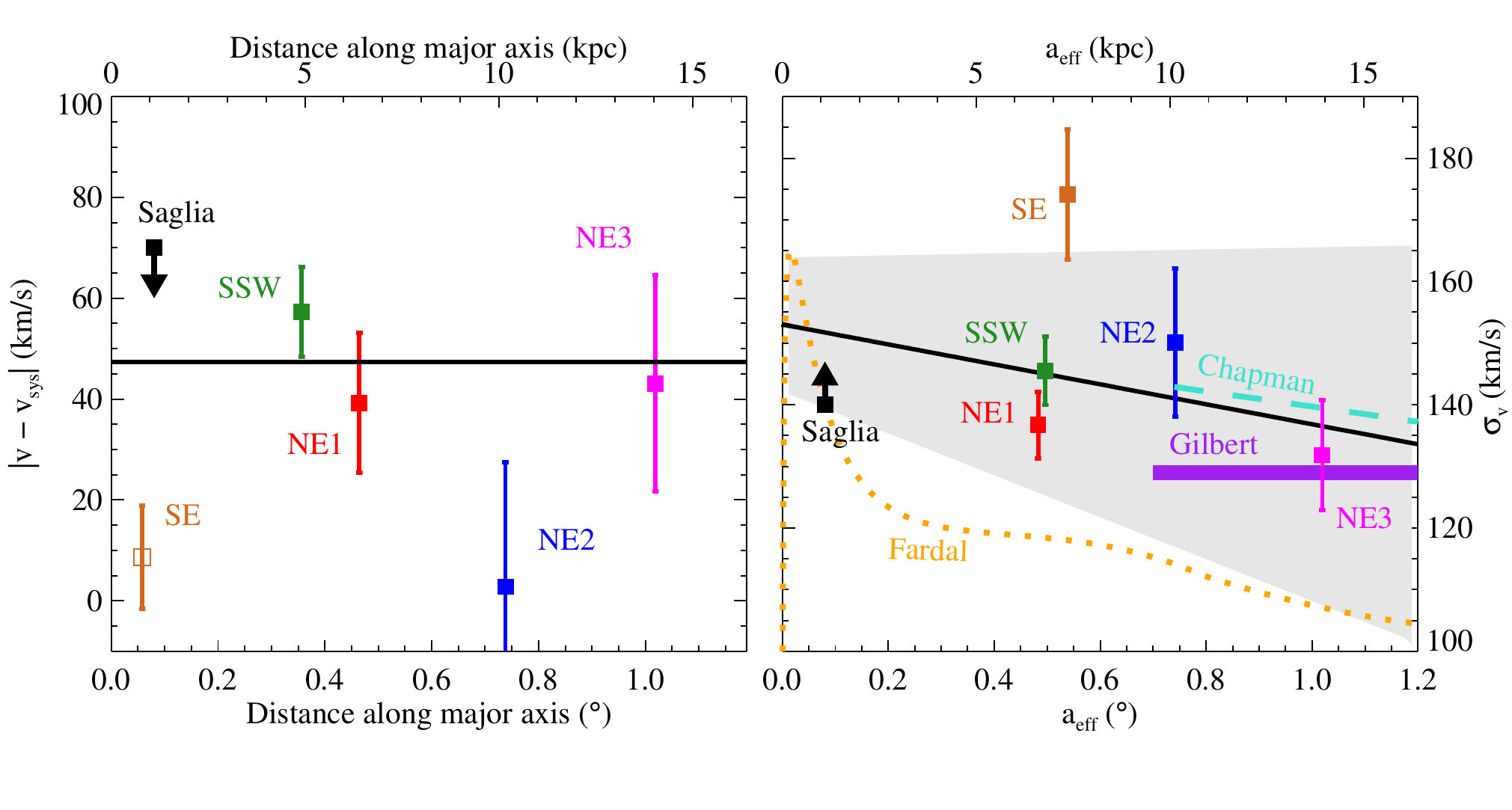}}
\centering
\caption{Velocity (left) and dispersion (right) profiles for M31's inner
  spheroid without accounting for the GSS or associated tidal
  debris. Left: Mean velocity is plotted against the projection along the major
  axis coordinate. The SE minor axis (brown open square) is close to
  the minor axis, so we do not expect to measure a significant ${|\overline{v}
    - v_{\rm sys}|}$ due to rotation. The black line in the left-hand plot is the mean
  value of the four off-minor-axis velocity measurements. Right: Dispersion
  is plotted versus $a_{\rm eff}$, the effective
  major axis coordinate for a spheroid with a 5:3 axis ratio. The best
  fit line to our dispersion measurements is shown by the solid black line, with the
  region within $\pm ~1\sigma$ of the best-fit line shaded in
  gray. Also plotted are limits on the bulge mean rotation velocity
  and velocity dispersion from \citet[black points with
  arrows]{sag10}, and velocity dispersion measurements further out in
  the spheroid by \protect\citet{cha06} and \protect\citet[horizontal purple
  line]{gil07}. The orange dotted line represents a model dispersion profile that includes isotropic bulge
  and halo stellar components, as well as a stellar disk and dark
  matter halo \protect\citep{far12}.} 
\label{fig_profiles_nogss}
\end{figure*}

The resulting fit for region NE1 is plotted in
Figure~\ref{fig_NE1_subs}. The first seven panels show the seven
subregions in NE1. In each panel, each walker at the end of 10,000 steps
is represented by a set
of colored lines: a red Gaussian with parameters $f_s,~
\barv_{\hot, r},~\sigma_{\hot, r}$ and a blue or green Gaussian
with parameters $(1-f_s), ~\barv_{\cold, s},~\sigma_{\cold, s}$. The
sum of these two is shown in violet. These distributions are overlaid
on the velocity histogram of stars in the subregion. The velocity ranges
excised for stream debris are shown in two ways: the velocity range
is shaded in light gray, and the stars in this range are colored red. 

The bottom middle panel shows the cumulative best-fit distributions:
the red line is the Gaussian corresponding to the mean
$(\barv_{\hot,r},~\sigma_{\hot,r})$. The blue curve is the cumulative
disk distribution: the sum of the
mean disk Gaussians from the seven subregions. The violet is the sum
of the other two curves. The excised velocity ranges are respresented
in the same way as in the subregion panels. 

The bottom right panel shows the positions of the walkers in
parameter space after 10,000 steps. There are eight concentrations of
points corresponding to the eight $(\barv,~\sigma_v)$ pairs (one spheroid
and seven disk). Each pair is summarized by an ellipse displaying the
mean and dispersion of that distribution.  

%Figure 6

%Figure 7

%FIGURE 8

\section{DISCUSSION}\label{sec_results}

This section is organized as follows. We first present the rotation curve
and velocity dispersion profile for the inner spheroid in
\S\,\ref{ssec_kin}. In \S\, \ref{ssec_cold} we
compare the velocity dispersion of the cold component to previous
measurements of the stellar disk as a sanity check on our analysis
method. In \S\,\ref{ssec_gss_results} we show that the exclusion of
velocity ranges corresponding to tidal debris from the GSS
significantly impacts the measured kinematical parameters of the
spheroid, but has minimal effect on the cold component. In
\S\,\ref{ssec_high_v} we compute a spheroid membership probability for
each star in our sample. In \S\,\ref{ssec_anisotropy} we explain that
the spheroid is likely supported by velocity anisotropy in addition to
rotation. Finally, in \S\,\ref{ssec_kstest} we show that neither
spectroscopic target selection criteria nor degree of crowding
introduces a significant bias towards the kinematically cold or hot
population. 

\subsection{Kinematical Parameters of the Inner Spheroid}\label{ssec_kin}

The spheroid distributions corresponding to the best-fit
$\overline{v}_{\rm sph}$ and $\sigma_{\rm sph}$ (hereafter $\sigma_v$) are
plotted in Figure~\ref{fig_6plots} for each of the five regions. (This
figure is simply a compilation of the summary panels of
Figures~\ref{fig_NE1_subs} and 13--16.) Kinematical and goodness-of-fit
parameters, accounting for the GSS and associated tidal debris,
are reported in Table~\ref{tab_results_gss}. The
$\chi^2$ probabilities have improved significantly from the uncorrected values
in Table~\ref{tab_results}. Four of the five regions now have
probabilities greater than 90\%, and that of the fifth (SSW) has
increased by a factor of 15.

\begin{figure*}[h]
\scalebox{1}{\includegraphics[trim=20 50 10 430, clip = true, width = 1\textwidth]{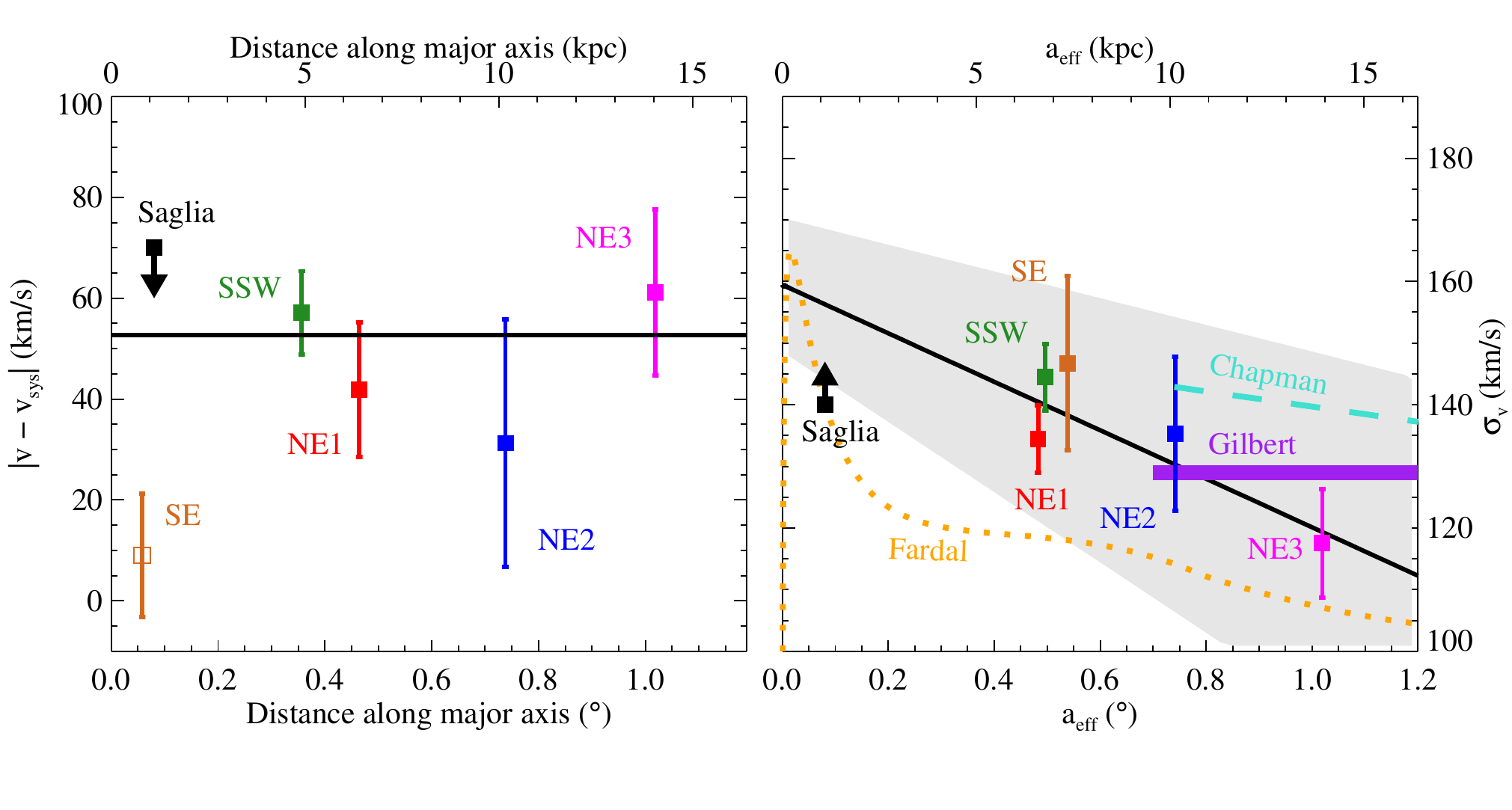}}
\centering
\caption{ Same as Figure~\ref{fig_profiles_nogss}, excluding stars in
  the velocity ranges corresponding to the GSS and associated tidal
  debris.}
\label{fig_profiles}
\end{figure*}

The dispersion profile is shown in the
right panel of Figure~\ref{fig_profiles}. The profile appears to
decrease smoothly with radius, though it is consistent with flat to
$2\sigma$. The best-fit line to this profile is

\begin{equation}
\sigma_v(a_{\rm eff})=(159.5 \pm 10.8) - (2.9 \pm
1.3)\frac{a_{\rm eff}}{1 ~\rm kpc}~\rm km~s^{-1}.
\label{eq_disp}
\end{equation}

\noindent where $a_{\rm eff}$ is the effective spheroid major axis
coordinate, assuming a 5:3 axis ratio \citep{pri94}. 
%We remind the reader
%that our fitting procedure makes use of the distinct velocity profiles
%of the individual subregions as shown in Figure~\ref{fig_NE1_subs},
%so that the fits are much better constrained than may be apparent from the
%combined distributions in Figure~\ref{fig_6plots}. 
Our dispersion profile is consistent with that measured by
\citet{gil07} but is slightly offset from other existing measurements,
including the
\citet{sag10} integrated-light measurement at 1.1 kpc on the major
axis of the bulge (black square in Figure~\ref{fig_profiles}) and the linear dispersion profile of
\citet{cha06}. These differences should be taken lightly, though,
because the measurements are not directly comparable. The
\citet{sag10} point may be slightly deflated by  contributions from
the low-dispersion disk stars, which those authors estimate
contribute about $30\%$ of the light in the slit. The \citet{cha06}
profile, meanwhile, is the innermost limit of measurements
primarily made farther out in the halo, so we do not necessarily expect
agreement at the radii covered by our study. 
%Chapman 06 -- radii R are projected radii; measurements are made
%mostly on the major axis. 

We also compare our results to the dispersion profile produced by a
model with spherical, isotropic, non-rotating bulge and halo stellar
components as well as a stellar disk and halo ($M/L =
2.5$) and bulge ($M/L = 5.6$) 
\citep[dotted orange line in Figure~\ref{fig_profiles};][]{far12}. The
model falls below our data at the $2\sigma$ level in the NE2 and
NE3 regions and at the $3\sigma$ level in the SE and NE1 regions. 
The mismatch suggests that, for example, the mass profile of the model
galaxy is too shallow, or the
gradient of the density profile of the tracer population is too
large. A more detailed analysis is beyond the scope of this paper.

Extracting the intrinsic rotation curve of the spheroid is nontrivial. The
relationship between the mean LOS velocity and the intrinsic rotation velocity
depends on the orbital dynamics of the spheroid, and in general is
difficult to determine without detailed 2D or 3D kinematical
mapping. Therefore, in the left
panel of Figure~\ref{fig_profiles} we simply present the mean LOS
component of the velocity versus the effective major axis
coordinate (based on a 5:3 axis ratio). Though two points are
consistent with zero to $2\sigma$, we detect significant rotation in
the SSW, NE1 and NE3 regions. The average value of $|v-v_{\rm M31}|$,
  excluding the SE minor axis point,  is $52.6 \pm 6.8~\rm km~s^{-1}$
  (solid line in Figure~\ref{fig_profiles}). All four off-minor axis
  points are consistent with this mean velocity to better than
  $1\sigma$. This is the first measurement of significant rotation in
  the inner spheroid. 

The velocity dispersion of M31's inner spheroid is
similar to that the halo of the MW at similar radii, but its mean velocity
is significantly larger. The
velocity ellipsoid of the MW's inner halo is $(\sigma_{V_R},
\sigma_{V_{\phi}}, \sigma_{V_Z})=(150\pm 2, 95\pm 2, 85\pm 1)$
\citep{car10}, on the same order as the LOS component of the M31
spheroid dispersion. However, the MW's inner halo has a mean rotation velocity
consistent with zero \citep{car10}.

%FIGURE 9
 
%FIGURE 10

\subsection{Dispersion and Velocity of the Cold Population}\label{ssec_cold}

The cold component $(\overline{v}, \sigma_v)$ for each subregion are plotted
against $r_{\rm deproj}$ in Figure~\ref{fig_disk}. As a sanity check
on our analysis method of fitting multiple Gaussians
to subregions, we compare these values to previously measured
kinematical parameters of the stellar disk. The cold component in a single
subregion has an average dispersion of $58 \rm~km~s^{-1}$,
reasonably consistent with the \citet{col11} dispersion measurements
of  $51~\rm km~s^{-1}$ and $36~\rm km~s^{-1}$ for the thick and thin
stellar disks, respectively. We expect our dispersion to be larger than the local
value, primarily because the finite spatial extent of our subregions
necessarily smears out the velocity distribution. For example, if
the spread in mean velocity due to $\Delta PA$ is $\squig 20~\rm
km~s^{-1}$ in a subregion, and the true velocity dispersion is
$~\squig 40~\rm km~s^{-1}$, then we would expect to measure an
effective dispersion of $\sqrt{20^2 + 40^2} = 45~\rm km~s^{-1}.$ This 
effect is accentuated if there is an additional spread in mean
velocity with radius. Therefore, the fact
that our cold component looks like the thick disk of \citet{col11}
does not imply that the thick component dominates the stellar disk.

%We also compare the
%mean velocity of the cold component in each subregion to the the
%expected disk rotation curve based on HI measurements as in
%\citet{cor10}. 

%At a given radius, deprojected to the plane of the
%disk, a cursory look demonstrates that the cold component 
%trails the gas with a lag on the order of $\sim 50~\rm km~s^{-1}$, similar
%to the $46~\rm km~s^{-1}$ velocity lag measured by \citet{col11}
%between the HI and thick stellar disks. Hence, we conclude that
%our minimal assumption of a locally cold disk is a reasonable way
%to account for the disk population. 

A possible concern is that contributions from multiple stellar disk components
(a combination of thin, thick, or extended) may invalidate our
assumption of a locally cold disk. If this were the case, we would
expect to see non-Gaussianity in the velocity signature of the cold
component. However, Table~\ref{tab_disk}
shows that that the velocity distributions of
the cold components in most of the subregions are well fit by a single
Gaussian. While this observation does not say anything about the
kinematical structure of the disk, it does suggest that our simple
assumption is adequate for our purpose of describing the spheroid. 

It is also possible that a kinematically warm thick disk component
would be incorporated into our Gaussian representation of the
spheroid; however, the high goodness-of-fit statistic again suggests
that this is not a significant effect. In any case, the bias on our
kinematical spheroid parameters induced by thick disk contamination
confirms, rather than invalidates, our qualitative results, as
discussed in \S\,\ref{ssec_anisotropy}.

A closer look at the best-fit cold and hot components by subregion in
Figures \ref{fig_NE1_subs} and 13--16 show that the trends in the cold component with
position angle match those expected for an inclined rotating disk. For example,
in Figure~\ref{fig_NE1_subs}, the mean velocity of the cold component
(blue and green curves) transitions from $-100~\rm km~s^{-1}$ in
subregion $\rm NE1_1$, along the major axis, to $-200~\rm km~s^{-1}$ in
subregion $\rm NE1_7$, farthest from the major axis. In other words,
the absolute value of the offset of the mean velocity of the cold
component from the systemic velocity of M31 moves closer to zero as we
march away from the major axis. This progression is clear in
Figures~\ref{fig_NE2_subs} and \ref{fig_NE3_subs} for regions NE2 and
NE3 as well, although it is less pronounced because these regions
subtend a smaller range in PA. 

\subsection{Effect of Tidal Debris Associated
  with the GSS on Spheroid Kinematics}\label{ssec_gss_results}

Accounting for the GSS and its associated debris has a significant effect on the
measured kinematical parameters of the underlying smooth inner
spheroid, supporting the
observation of \citet{far07} that the debris was visible in the samples
of \citet{iba05}, \citet{cha06}, and \citet{mer06}.

It is unclear whether the secondary stream identified by \citet{kal06b}
and confirmed by \citet{gil09} is present in the NE1--NE3
regions. Figure~\ref{fig_v_rproj_all} reveals a second clump of
objects in the NE2 region, offset from the primary stream by
75--100$~\rm km~s^{-1}$, the same separation as that between the two
streams south of the galaxy. However, such clumps are barely, if at
all, visible in the NE1 and NE3 regions, and their exclusion does not
significantly affect the kinematical parameters of the spheroid. Further
observations are necessary to determine if there is a second
substructure and, if so, whether it has a physical connection to the GSS.

Is the GSS the only source of nonvirialized
substructure biasing our measurements of the mean velocity and
dispersion of the inner spheroid? While we cannot prove that all stars
except those in the excised velocity ranges belong to either a disk or
a perfectly smooth spheroid, we can show that the effect of other
nonvirialized substructure on our kinematical characterization of the
underlying smooth spheriod is negligible. In star-count maps of the
inner regions of M31, such as those in \citet{iba01}, the GSS
is by far the most prominent substructure. Even so, GSS stars account
for only a small fraction of our spectroscopic sample (see pink shaded
regions in Figure~\ref{fig_6plots}), and have a relatively small effect
on our results. Other, less-prominent substructures would (1) be
nearly impossible to detect and account for, and (2) have a negligible
effect on the measured kinematical parameters of the inner spheroid. 

\subsection{Spheroid/Disk Membership Probability and Extreme Velocity
  Stars}\label{ssec_high_v}

\begin{figure}[]
\scalebox{1}{\includegraphics[trim=20 0 175 315, clip = true, width = 0.48\textwidth]{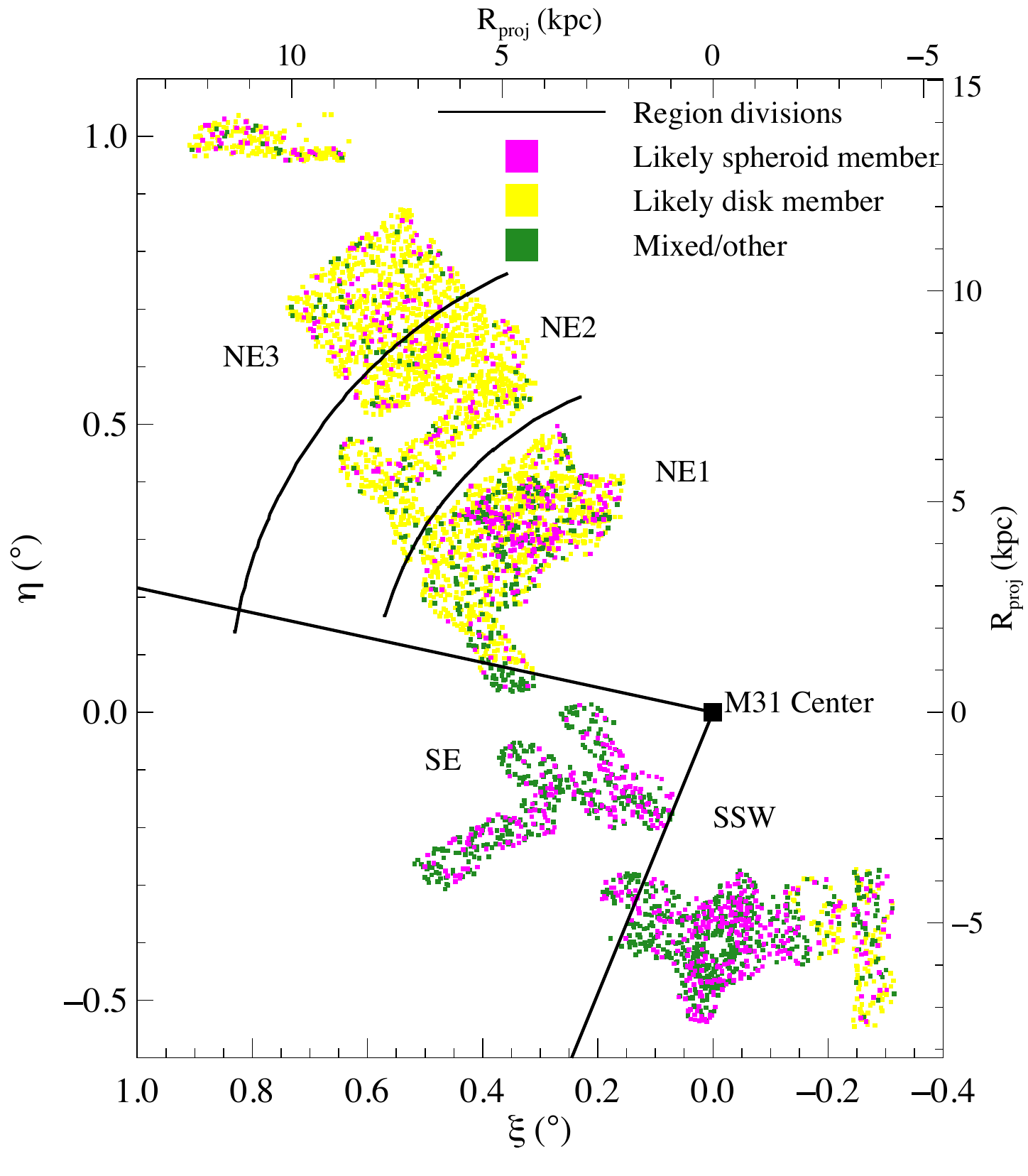}}
\centering
\caption{Probable subcomponent membership of each star based on
  location and velocity. Stars at least 3 times more likely to be a disk
  member than anything else are shown in yellow;
  those at least 3 times
  more likely than anything else to be a spheroid member are shown in
  pink; all other objects, including likely M32
  members, are shown in green. Region divisions are shown as solid lines.}
\label{fig_bulge_prob}
\end{figure}

We can apply our subregion fits to quantify the spheroid
membership probability for any star. At the location and velocity of
each star in our sample, we calculate the ratio of the values of the best-fit hot
component in that region and the disk and M32 components in that
subregion. The disk and M32 membership likelihoods are
calculated in a similar fashion. The results are sorted into three categories as
follows: stars that are at least three times as likely to be disk
members as anything else (yellow in Figure~\ref{fig_bulge_prob}); stars
at least three times as likely to be spheroid members as anything else
(magenta); and other objects, including likely M32 members (green). As
expected, objects on the major axis are much more likely to be disk
members than are those on the minor axis. Most important, we see
likely spheroid members at all radii covered by our sample. 

Recently, \citet{cal10} reported discovery of an ``extreme velocity''
star at a projected radius of $4$ kpc ($0.3^{\circ}$) along the SW major
axis of M31. The star has a velocity of $-780~\rm km~s^{-1}$, essentially
excluding it from membership in the thin, cold stellar disk. Those
authors attribute the star's highly negative velocity to possible
membership in the GSS, even though it would have to be a $6\sigma$
outlier of the stream velocity distribution. Our probability map
demonstrates that this object can more easily be interpreted as a
member of the spheroid, even at these large radii: it may be a
$4\sigma$ outlier of the spheroid distribution. 

It is tempting to interpret this distribution of probabilities as a map of
bulge-to-disk fraction. However, this statistic can be more
reliably constrained using photometric light-profile fitting in
conjunction with kinematical decomposition. This study will be
presented in Dorman et~al. (2012, in prep).

%FIGURE 11

%FIGURE 12
\begin{figure}[]
\scalebox{1}{\includegraphics[trim = 10 45 50 270, clip =  true, width
  = 0.48\textwidth]{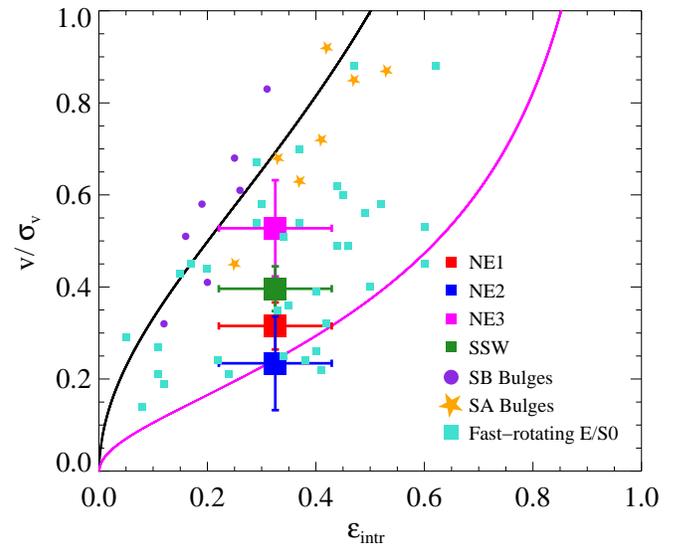}}
\centering
\caption{Anisotropy diagram for spheroidal systems. The large filled
  squares show the ratio of the observed mean velocity and the
  velocity dispersion against the intrinsic ellipticity
  of M31's spheroid. Error bars represent the range of
  reported ellipticities converted to edge-on intrinsic ellipticities
  as explained in \protect\citet{cap07}. The black curve shows the expected
  value of the ratio $\overline{v}/\sigma_v$ as a function of ellipticity for an isotropic,
  rotating galaxy. The magenta line approximates the behavior of an
  oblate anisotropic galaxy with anisotropy parameter $\delta \sim 0.7
  \epsilon$ \protect\citep{cap07}. Also shown are fast-rotating E/S0 galaxies
  from SAURON (turquoise squares) and bulges of barred [violet
  circles, \protect\citep{kor82b}] and unbarred [orange stars, \protect\citep{kor82}] spirals. The
  inner spheroid of M31 appears to be flattened by velocity anisotropy
  in addition to rotation, and therefore more closely resembles an
elliptical galaxy than a spiral galaxy bulge.}
\label{fig_ellipticity}
\end{figure}

\subsection{Anisotropy}\label{ssec_anisotropy}
We investigate the degree to which the flattening of the spheroid may
be due to rotation by comparing the spheroid ellipticity $\epsilon=1-b/a$ to
the ratio $v/\sigma_v$. The value of $\epsilon$ is
uncertain at these radii. \citet{pri94} measure $\epsilon=0.4$ with
limited data at $R_{\rm proj}=10~\rm kpc$. \citet{cou11} perform a fit
to a more extended data set to obtain values between $0.21$ and $0.37$,
depending on their bulge/disk decomposition and modeling technique,
for a relatively small bulge with scale length $\sim1.0~\rm kpc$.

Despite this range of possible ellipticities,  $\overline{v}/\sigma_v$ of the
spheroid in every region is lower than that of a rotationally
flattened oblate isotropic rotator (black line in
Figure~\ref{fig_ellipticity}). We measure $\overline{v}/\sigma_v$ = 0.23--0.52,
but $v_{\rm  rot}/\sigma_v$ of $0.87$ or $0.54$ would be required for rotation
to flatten the spheroid to an ellipticity of $0.4$ or $0.21$,
respectively. Hence, rotation alone probably does not account for all
of the flattening of the spheroid. 

It is possible that anisotropy in the velocity ellipsoid can provide
the remainder of the flattening. Anisotropy can be parameterized
by 

\begin{equation}
\delta \equiv 1 - \frac{\Pi_{zz}}{\Pi_{xx}}
\end{equation}

\noindent where $z$ points along the axis of symmetry of the spheroid, $x$ is
any orthogonal direction (say, $r$), and $\Pi_{kk}$ represents the
pressure from velocity dispersion along direction $k$
\citep[e.g.][]{cap07}. In a system with an axisymmetric velocity ellipsoid,
$\delta=0$ corresponds to a system whose flattening is unaffected by
anisotropy, while  $\delta \sim 1$ corresponds to a system whose
flattening is almost entirely due to anisotropy. Using the iterative method described in \citet{cap07}, we find that an
anisotropy of $\delta=$~0.05--0.27 is required to explain a spheroid
ellipticity of 0.21--0.4, given the mean value of 
$\overline{v}/\sigma_v$ in regions NE1--NE3 and SSW. 

We compare the inner spheroid of M31 to other
spheroidal systems. The bulges
of spiral galaxies generally fall on or above the line for
rotationally-flattened systems \citep{kor82, kor82b}. In contrast, 
so-called ``fast-rotating'' ellipticals tend to lie between this
line and the line $\delta \sim 0.7\epsilon_{\rm intr}$ (magenta in
Figure~\ref{fig_ellipticity}) \citep{cap07}. However, these comparison
systems are observed at around one effective radius
$(r_{\rm eff})$, whereas our kinematical measurements range from
1.5--14$~r_{\rm eff}$. The inner spheroid of M31, then, more closely
resembles the inner $r_{\rm eff}$ of a fast-rotating elliptical than
the central bulge of a spiral galaxy. 

As suggested earlier in \S\,\ref{ssec_cold}, it is possible that  the
spheroid velocity distribution may be contaminated by thick disk
stars. If so, the true spheroid mean velocity is lower than we report,
and the true spheroid dispersion higher. However, note that the
$\overline{v}/\sigma_v$ of the spheroid is already small enough to look more like
an elliptical galaxy than the bulge of a spiral galaxy. The possible
bias induced by a thick disk component would merely increase this
effect, confirming our conclusion that M31's inner spheroid rotates
unusually slowly.

\subsection{Effect of Target Selection Criteria on Velocity
  Distribution}\label{ssec_kstest}
The majority of our spectroscopic targets were chosen on the basis of
$i^{\prime}$ magnitude only, but a small fraction of the targets in
the MCT region were identified based on position in the PHAT CMD. This
latter category contains proportionally more of the rarer metal-poor and
metal-intermediate RGB populations. If metal-poor and
metal-intermediate RGBs preferentially trace the kinematically hot
population, then inclusion of the PHAT-selected objects could bias our
measurements, especially our spheroid membership probabilities. To
test for bias, we performed a Komolgorov-Smirnov (K-S) test comparing the
velocity distributions of the PHAT-selected targets
and the magnitude-selected targets in each subregion. The test confirmed
that the two distributions were indistinguishable. 

Similarly, we used a K-S test to confirm that inclusion of crowded
objects does not bias us towards one structural subcomponent. We
created two lists of objects: ``crowded'' (those sharing a slit with
at least one serendipitously detected neighbor) and ``isolated'' (those
without any such neighbors). Again, there was no singificant
difference between the velocity distributions of the two categories in
any subregion.

% TABLE 2 - DONE 
\begin{deluxetable*}{lllccccccccccc}
\tabletypesize{\scriptsize}
\tablecaption{Uncorrected Kinematical Spheroid Parameters}
\tablewidth{0 pt}
\tablehead{
\colhead{Region}&
\colhead{$\xi~(^\circ)$}&
\colhead{$\eta~(^\circ)$}&
\colhead{$R_{\rm proj}$ (kpc)}&
\colhead{$a_{\rm eff}$ (kpc)}&
\colhead{$N_{\rm region}$}&
\colhead{$\overline{v} - \overline{v}_{M31}~(\rm km~s^{-1})$}&
\colhead{$\sigma_v~(\rm km~s^{-1})$}&
\colhead{$N_{\rm spheroid}/N_{\rm total}$}&
\colhead{$\chi^2$ prob}&
}
\startdata

\vspace{2mm}
NE1 & ~~0.40 & ~~0.31 & ~6.45 & ~6.81 & 1615 & $+$39.3 $\pm$ 13.9 &
136.7 $\pm$ ~5.4 & 0.288 & 0.731\\
\vspace{2mm}
NE2 & ~~0.48 & ~~0.55 & 10.11 & 10.31 & 1180 & ~$+$2.8 $\pm$ 24.6 &
150.1 $\pm$ 12.0 & 0.120 & 0.136 \\
\vspace{2mm}
NE3 & ~~0.64 & ~~0.79 & 13.93 & 14.00 & ~859 & $+$48.0 $\pm$ 24.0 &
131.4 $\pm$ 9.9 & 0.203 & 0.930 \\
\vspace{2mm}
SE & ~~0.26 & $-$0.14 & ~4.30 & ~6.91 & ~684 & $+$8.7 $\pm$ 10.3  &
174.1 $\pm$ ~10.5  & 0.522 & 0.121\\
\vspace{2mm}
SSW & $-$0.09 & $-$0.40 & ~5.69 & ~6.73 & 1313 & $-$57.3 $\pm$~8.9 &
145.5 $\pm$ ~5.5 & 0.465 & 1.5$\times10^{-5}$ \\

\tableline \vspace{-3mm}
\enddata
\label{tab_results}
\end{deluxetable*}

%TABLE 3 - DONE 
\begin{deluxetable*}{lllccccccccc}
\tabletypesize{\scriptsize}
\tablecaption{Kinematical Spheroid Parameters Corrected for
Tidal Debris}
\tablewidth{0 pt}
\tablehead{
\colhead{Region}&
\colhead{$\eta~(^\circ)$}&
\colhead{$\xi~(^\circ)$}&
\colhead{$R_{\rm proj}$ (kpc)}&
\colhead{$a_{\rm eff}$ (kpc)}&
\colhead{$N_{\rm region}$}&
\colhead{$\overline{v} - \overline{v}_{M31}~(\rm km~s^{-1})$}&
\colhead{$\sigma_v~(\rm km~s^{-1})$}&
\colhead{$N_{\rm spheroid}/N_{\rm total}$}&
\colhead{$\chi^2$ prob}&
}
\startdata

\vspace{2mm}
NE1 & ~~0.40 & ~~0.31 & ~6.45 & ~6.81 & 1615 & $+$41.8 $\pm$ 13.3 &
134.4 $\pm$ ~5.5 & 0.297 & 0.945 \\
\vspace{2mm}
NE2 & ~~0.48 & ~~0.55 & 10.11 & 10.31 & 1180 & ~$+$31.3 $\pm$ 24.5 &
135.3 $\pm$ 12.5 & 0.130 & 1.000 \\
\vspace{2mm}
NE3 & ~~0.64 & ~~0.79 & 13.93 & 14.00 & ~859 & $+$61.2 $\pm$ 16.44 &
117.5 $\pm$ 8.8 & 0.198 & 0.998 \\
\vspace{2mm}
SE & ~~0.26 & $-$0.14 & ~4.30 & ~6.91 & ~684 & $+$9.0 $\pm$ 12.2  &
144.5 $\pm$ ~5.5  & 0.571 & 0.819\\
\vspace{2mm}
SSW & $-$0.09 & $-$0.40 & ~5.69 & ~6.73 & 1313 & $-$57.1 $\pm$~8.2 &
145.4 $\pm$ ~6.5 & 0.484 & 2.1$\times10^{-4}$ \\

\tableline \vspace{-3mm}
\enddata
\label{tab_results_gss}
\end{deluxetable*}

 %-----------------------------------------------------------
 %  6. SUMMARY & CONCLUSIONS
 %-----------------------------------------------------------
\section{SUMMARY \& CONCLUSIONS}\label{sec_summary}
We have measured reliable radial velocities of over five thousand red giant
branch stars in the inner $20 \rm ~kpc$ of M31 using the Keck/DEIMOS
multiobject spectrograph. Targets were selected using a series of
statistical tests designed to identify isolated M31 members bright
enough to yield quality spectra. 

By fitting a locally cold disk and a kinematically hot spheroid to the
velocity distribution with an MCMC algorithm, we have measured the most probable kinematical parameters
$\overline{v}$ and $\sigma_v$ of the red giant stellar population of the inner
spheroid of M31 in each of five spatial bins. We find that, though the
raw values are inconsistent with a physical rotation pattern,
accounting for the presence of tidal debris due to the
GSS
allows us to detect a significant spheroid rotation velocity. 

We find that probable spheroid members are present at all radii in our
sample. When used in conjunction with integrated-light measurements, these membership
probabilities will be a powerful tool for rigorous bulge/disk decomposition in the inner parts of M31. 

We also compare the $\overline{v}/\sigma_v$ of the inner spheroid to
those of other spheroidal structures. We find that rotation is
insufficient to explain the flattening of the inner spheroid; it more
closely resembles an anisotropic fast-rotating elliptical than the bulge of
a spiral or barred spiral galaxy as measured at $\sim 1~r_{\rm eff}$. 

Our magnitude-limited survey biases us towards the bright, old stellar
population. As more PHAT data becomes available over the next few years,
we will be able to select spectroscopic targets that represent a range
of stellar populations within our magnitude range. We plan to present
an analysis of the kinematics of different stellar populations in an
upcoming paper. 

%-----------------------------------------------------------

 %  ACKNOWLEDGEMENTS
 %-----------------------------------------------------------
\section{Acknowledgments}

We would like to thank the referee Alan McConnachie for helpful
comments and Roeland van der Marel and Elisa Toloba for useful
discussions. 

PG, KH, MF and CD acknowledge NSF grants AST-0607852 and
AST-1010039. PG, KH and CD acknowledge NASA grant
HST-GO-12055. Additionally, CD was supported by a University of California Eugene
Cota-Robles Graduate Fellowship and a NSF Graduate Research
Fellowship. 

We wish to recognize and acknowledge the very significant
cultural role and reverence that the summit of Mauna Kea has always
had within the indigenous Hawaiian community. We are most fortunate to
have the opportunity to conduct observations from this mountain. 

%% Make sure that you provide a unique KEY for every \bibitem or else they



%\clearpage
%\bibliography{spheroid}

\begin{thebibliography}{51}%\expandafter\ifx\csname
                           %natexlab\endcsname\relax\def\natexlab#1{#1}\fi

\expandafter\ifx\csname natexlab\endcsname\relax\def\natexlab#1{#1}\fi
\bibitem[{{Athanassoula \& Beaton}(2006)}]{ath06}
{Athanassoula}, E. \& {Beaton}, R.~L. 2006, \mnras, 370, 1499

\bibitem[{{Beaton} {et~al.}(2007)}]{bea07}
{Beaton}, R.~L., {Majewski}, S.~R., {Guhathakurta}, P., et~al. 2007,
\apj, 658, L91

%\bibitem[{{Beers} {et~al.}(2011)}]{bee11}
%{Beers}, T.~C., {Carollo}, D., {Ivezic}, Z., et~al. 2011, submitted
%(arXiv:1104.2513B)

\bibitem[{{Brown} {et~al.}(2006)}]{bro06}
{Brown}, T.~M., {Smith}, E., {Ferguson}, H.~C., et~al. 2006, \apj,
652, 323

\bibitem[{{Bullock \& Johnston}(2005)}]{bul05}
{Bullock}, J.~S. \& {Johnston}, K.~V. 2005, \apj, 635, 931 

\bibitem[{{Caldwell} {et~al.}(2010)}]{cal10}
{Caldwell}, N., {Morrison}, H., {Kenyon}, S.~J., et~al. 2010, \apj,
139, 372

\bibitem[{{Cappellari} {et~al.}(2007)}]{cap07}
{Cappellari}, M., {Emsellem}, E., {Bacon}, R., et~al. 2007, \mnras, 379, 418

\bibitem[{{Carollo} {et~al.}(2007)}]{car07}
{Carollo}, D., {Beers}, T.~C., {Lee}, Y.~S., et~al. 2007,
Nature, 450, 1020

\bibitem[{{Carollo} {et~al.}(2010)}]{car10}
{Carollo}, D., {Beers}, T.~C., {Chiba}, M., et~al. 2010, \apj, 712, 692

\bibitem[{{Chapman} {et~al.}(2006)}]{cha06}
{Chapman}, S.~C., {Ibata}, R., {Lewis}, G.~F., et~al. 2006, \apj, 653, 255

\bibitem[{{Choi} {et~al.}(2002)}]{cho02}
{Choi}, P.~I., {Guhathakurta}, P., \& {Johnston}, K.~V. 2002, \aj,
124, 310

\bibitem[{{Collins} {et~al.}(2011)}]{col11}
{Collins}, M.~L.~M., {Chapman}, S.~C., {Ibata}, R.~A., et~al. 2011,
\mnras, 413, 1548

\bibitem[{{Corbelli} {et~al.}(2010)}]{cor10}
{Corbelli}, E., {Lorenzoni}, S., {Walterbos}, R., {Braun}, R. \&
{Thilker}, D. 2010, A\&A, 511, A89

\bibitem[{{Courteau}(1996)}]{cou96}
{Courteau}, S. 1996, \apjs, 103, 363

\bibitem[{{Courteau} {et~al.}(2011)}]{cou11}
{Courteau}, S., {Widrow}, L.~M., {McDonald}, M., et~al. 2011,
\apj, in press (arXiv: 1106.3564)

\bibitem[{{Dalcanton} {et~al.}(2012)}]{dal12}
{Dalcanton}, J.~J., {Williams}, B.~F., {Lang}, D., et~al. 2012, \apjs
(arXiv: 1204.0010)

\bibitem[{{Davis} {et~al.}(2003)}]{dav03}
{Davis}, M., {Faber}, S.~M., {Newman}, J., et~al. 2003, SPIE, 4834,
161

\bibitem[{{Fardal} {et~al.}(2007)}]{far07}
{Fardal}, M.~A., {Guhathakurta}, P., {Babul}, A., \& {McConnachie},
A.~W. 2007, \mnras, 380, 15

\bibitem[{{Fardal} {et~al.}(2012)}]{far12}
{Fardal}, M.~A., {Guhathakurta}, P., {Gilbert}, K.~M., et~al. 2012, \mnras, submitted

\bibitem[{{Foreman-Mackey} {et~al.}(2012)}]{for12}
{Foreman-Mackey}, D., {Hogg}, D.~W., {Lang}, D. \& {Goodman}, J. 2012
(arXiv: 1292.3665)

\bibitem[{{Geha} {et~al.}(2006)}]{geh06}
{Geha}, M., {Guhathakurta}, P., {Rich}, R.~M. \& {Cooper}, M.~C. 2006,
\aj, 131, 332

\bibitem[{{Geha} {et~al.}(2010)}]{geh10}
{Geha}, M., {van~der~Marel}, R.~P., {Guhathakurta}, P., et~al. 2010,
\apj, 711, 361

\bibitem[{{Gilbert} {et~al.}(2007)}]{gil07}
{Gilbert}, K.~M., {Guhathakurta}, P., {Kalirai}, J.~S., et~al. 2007,
\apj, 652, 1188	

\bibitem[{{Gilbert} {et~al.}(2009)}]{gil09}
{Gilbert}, K.~M., {Guhathakurta}, P., {Kollipara}, P., et~al. 2009,
\apj, 705, 1275

\bibitem[{{Goodman \& Weare}(2010)}]{gw10}
{Goodman}, J. \& {Weare}, J., 2010, Comm.~App.~Math.~Comp.~Sci., 5(1),
65

\bibitem[{{Guhathakurta} {et~al.}(1988)}]{guh88}
{Guhathakurta}, P., {van Gorkom}, J.~H., {Kotanyi}, C.~G. \&
{Balkowski}, C. 1988, \aj, 96, 3

\bibitem[{{Guhathakurta} {et~al.}(2005)}]{guh05}
{Guhathakurta}, P., {Osteimer}, J.~C., {Gilbert}, K.~M., et~al. 2005
(arXiv: astro-ph/050236v5)

\bibitem[{{Guhathakurta} {et~al.}(2006)}]{guh06}
{Guhathakurta}, P., {Rich}, R.~M., {Reitzel}, D.~B., et~al. 2006,
\apj, 131, 2497

\bibitem[{{Howley} {et~al.}(2008)}]{how08}
{Howley}, K.~M., {Geha}, M., {Guhathakurta}, P., et~al. 2008, \apj,
683, 722

\bibitem[{{Howley} {et~al.}(2012)}]{how12}
{Howley}, K.~M., {Guhathakurta}, P., {van der Marel}, R., et~al. 2012,
\apj, submitted (arXiv: astro-ph/1202.2897)


\bibitem[{{Hubble}(1936)}]{hubble}
{Hubble}, E.~P. 1936, The Realm of the Nebulae (New Haven: Yale
Univ. Press)

\bibitem[{{Ibata} {et~al.}(2001)}]{iba01}
{Ibata}, R., {Irwin}, M., {Lewis}, G., {Ferguson}, A.~M.~N. \&
{Tanvir}, N. 2001, Nature, 412, 49I

\bibitem[{{Ibata} {et~al.}(2005)}]{iba05}
{Ibata}, R., {Chapman}, S., {Ferguson}, A.~M.~N, et~al. 2005, \apj,
634, 

\bibitem[{{Ibata} {et~al.}(2007)}]{iba07}
{Ibata}, R., {Marin}, N.~F., {Irwin}, M., {Chapman}, S., et~al. 2007,
\apj, 634, 287

\bibitem[{{Irwin} {et~al.}(2005)}]{irw05}
{Irwin}, M.~J., {Ferguson}, A.~M.~N., {Ibata}, R.~A., {Lewis}, G.~F. \&
 {Tanvir}, N.~R. 2005, \apj, 628, L105

%kal06a is the metallicity paper
\bibitem[{{Kalirai} {et~al.}(2006a)}]{kal06a}
{Kalirai}, J.~S., {Gilbert}, K.~M., {Guhathakurta}, P., et~al. 2006, \apj, 648, 389

%kal06b is the secondary stream paper
\bibitem[{{Kalirai} {et~al.}(2006b)}]{kal06b}
{Kalirai}, J.~S., {Guhathakurta}, P., {Gilbert}, K.~M., et~al. 2006,
\apj, 641, 268

%Disk galaxies
\bibitem[{{Kormendy \& Illingworth}(1982)}]{kor82}
{Kormendy}, J. \& {Illingworth}, G. 1982, \apj, 256, 460

%Barred galaxies
\bibitem[{{Kormendy}(1982b)}]{kor82b}
{Kormendy}, J. 1982, \apj, 257, 75

\bibitem[{{Kormendy \& Kennicutt}(2004)}]{kor04}
{Kormendy}, J. \& {Kennicutt}, R.~C. 2004,  Annual Reviews of
Astronomy \& Astrophysics, 42, 603

\bibitem[{{McConnachie} {et~al.}(2008)}]{mcc05}
{McConnachie}, A.~W., {Irwin}, M.~J., {Ferguson}, A.~M.~N., {Ibata},
R.~A., {Lewis}, G.~F. \& {Tanvir}, N. 2005, \mnras, 356, 979

\bibitem[{{Merrett} {et~al.}(2006)}]{mer06}
{Merrett}, H.~R., {Merrifield}, M.~R., {Douglas}, N.~G., {et~al} 2006,
\mnras, 369, 120

\bibitem[{{Peng} {et~al.}(2002)}]{pen02}
{Peng}, C.~Y., {Ho}, L.C., {Impey}, C.~D. \& {Rix}, H. 2002, \aj, 124,
266

\bibitem[{{Pritchet \& van den Bergh}(1994)}]{pri94}
{Pritchet}, C.~J., \& {van den Bergh}, S. 1994, \aj, 107, 1730

\bibitem[{{Reitzel} {et~al.}(1998)}]{rei98}
{Reitzel}, D.~B., {Guhathakurta}, P. \& {Gould}, A. 1998, \aj, 116,
707

\bibitem[{{Saglia} {et~al.}(2010)}]{sag10}
{Saglia}, R.~P., {Fabricius}, M., {Bender}, R., et~al. 2010, A\&A, 509, A61

\bibitem[{{Simard}(2002)}]{sim02}
{Simard}, L. 2002, \apjs, 142, 1

\bibitem[{{Simon \& Geha}(2007)}]{sim07}
{Simon}, J.~D. \& {Geha}, M. 2007, \apj, 670, 313

\bibitem[{{Sohn} {et~al.}(2007)}]{soh07}
{Sohn}, S.~T., {Majewski}, S.~R., {Mu{\~n}oz}, et~al. 2007, \apj,
663, 960

\bibitem[{{Stetson}(1994)}]{ste94}
{Stetson}, P.~B. 1994, PASP, 106, 250

\bibitem[{{Tollerud}(2011)}]{tol11}
{Tollerud}, E.~J., {Beaton}, R.~L., {Geha}, M.~C., et~al. 2011, \apj,
submitted (arXiv: 1112:1067)

\bibitem[{{Zolotov} {et~al.}(2010)}]{zol10}
{Zolotov}, A., {Willman}, B., {Brooks}, A.~M., et~al. 2010, \apj, 721, 738

\end{thebibliography}

%%%%%%%%FIGURES%%%%%%%%%

\clearpage
\appendix
\section{MCMC results for regions NE2, NE3, SE, and SSW}

Figures 13--16 are the analogs of Figure~\ref{fig_NE1_subs} for
regions NE2, NE3, SE, and SSW, respectively. In this appendix, we briefly describe
the results in each of these regions. 

\subsection{NE2 Region}
\begin{figure*}[h]
\scalebox{1.0}{\includegraphics[trim = 15 5 55 360, clip =  true,
  width = 1\textwidth]{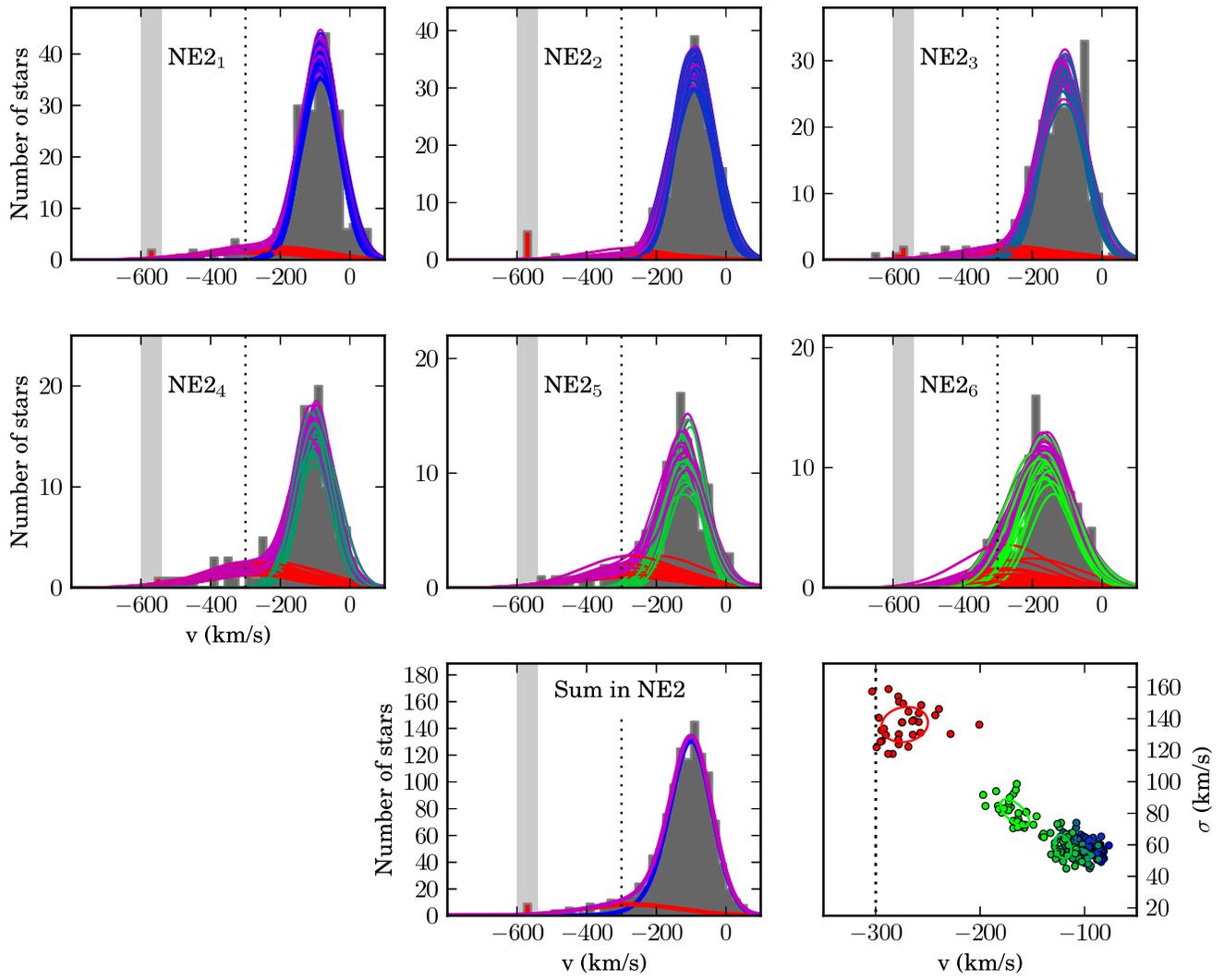}}
\caption{Same as Figure~\ref{fig_NE1_subs}, but for the NE2 region.}
\label{fig_NE2_subs}
\end{figure*}

Eleven stars are
excluded from the MCMC fit in the NE2 region due to possible tidal
debris membership; as seen in Figure~\ref{fig_NE2_subs}, all
are from the four subregions closest to the major axis. Exclusion drastically improves the
chi-squared probability that the spheroid plus disk Gaussians are a good
representation of the velocity distribution, raising it from 0.136 to
1.000. Exclusion also increases the mean velocity and decreases the
velocity dispersion of the spheroid, bringing them much more into
agreement with those measured in the other regions. 

Several trends characterize the subregion panels. First, the fraction
of stars in the spheroid (parameterized by ratio of the areas of the
red spheroid and blue/green disk curves) increases with increasing
distance from the major axis. Second, the mean velocity of the cold
component decreases with increasing distance from the major
axis. Reading off the velocity axis of the parameter-space view in the
bottom right panel, we see that the mean velocity of the cold
component moves from about $-90$ to $-170~\rm km~s^{-1}$. 
The minimum velocity is less negative than that in region NE1 because region
NE2 subtends a smaller range in PA. 

\subsection{NE3 Region}
Nine stars are excluded from region NE3 as possible tidal debris
contaminants. Exclusion slightly improves the reduced chi-squared of
the fit. It also slightly increases the mean velocity of the hot component
and decreases the velocity dispersion, bringing them more into
agreement with those measured in the other regions. 

\begin{figure*}[h]
\scalebox{1.0}{\includegraphics[trim = 40 145 55 360, clip =  true,
  width = 1\textwidth]{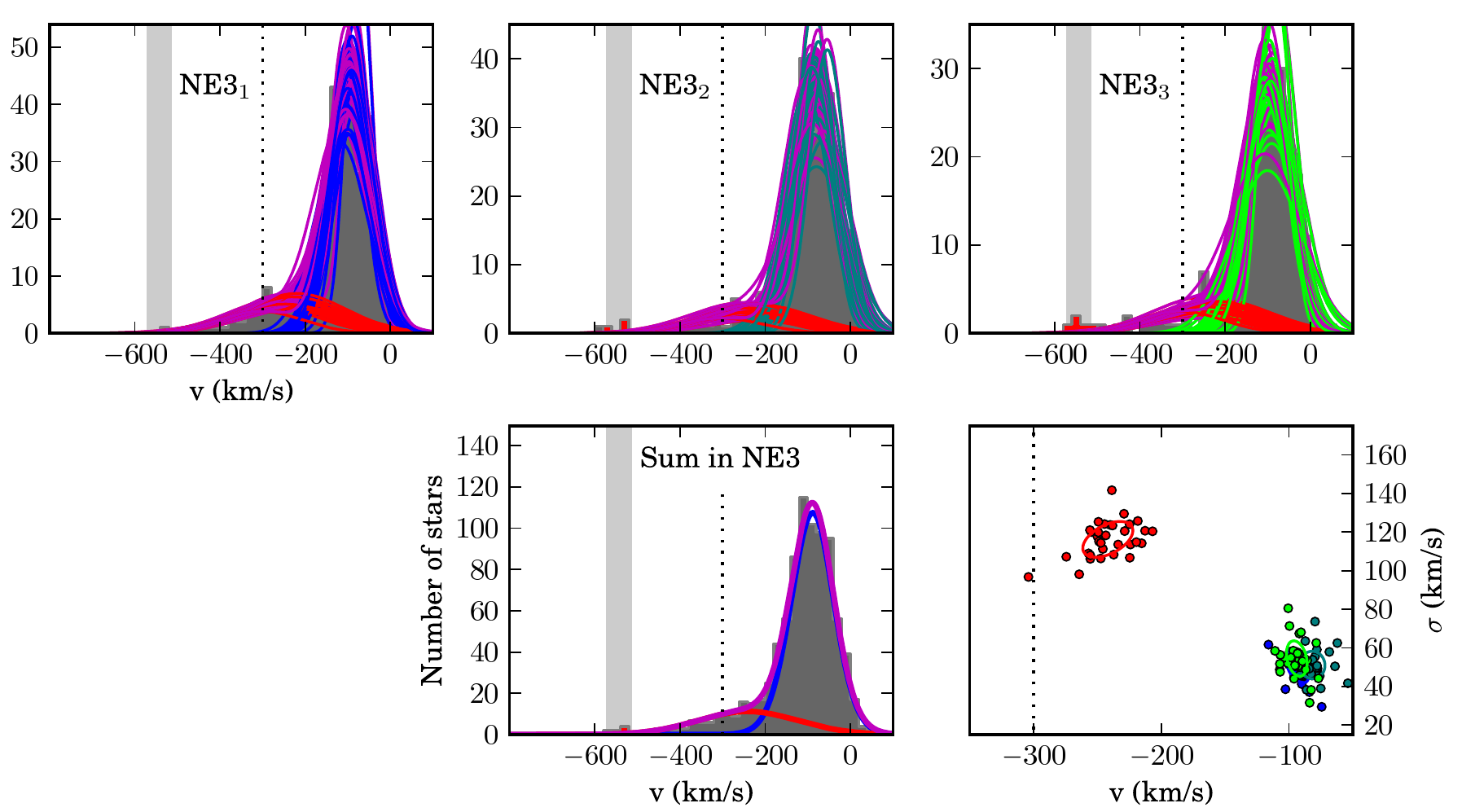}}
\centering
\caption{Same as Figure~\ref{fig_NE1_subs}, but for the NE3 region.}
\label{fig_NE3_subs}
\end{figure*}

Because this region is centered farthest from the center of M31, it
subtends the smallest range in PA of the three NE regions, and so the
variation in cold component mean velocity between subregions is
small. However, the parameter-space view in
Figure~\ref{fig_NE3_subs} shows that the cold component kinematical
parameters are very well defined. 

\subsection{SE Region}
\begin{figure*}[h]
\scalebox{1.0}{\includegraphics[trim = 40 145 55 360, clip =  true,
  width = 1\textwidth]{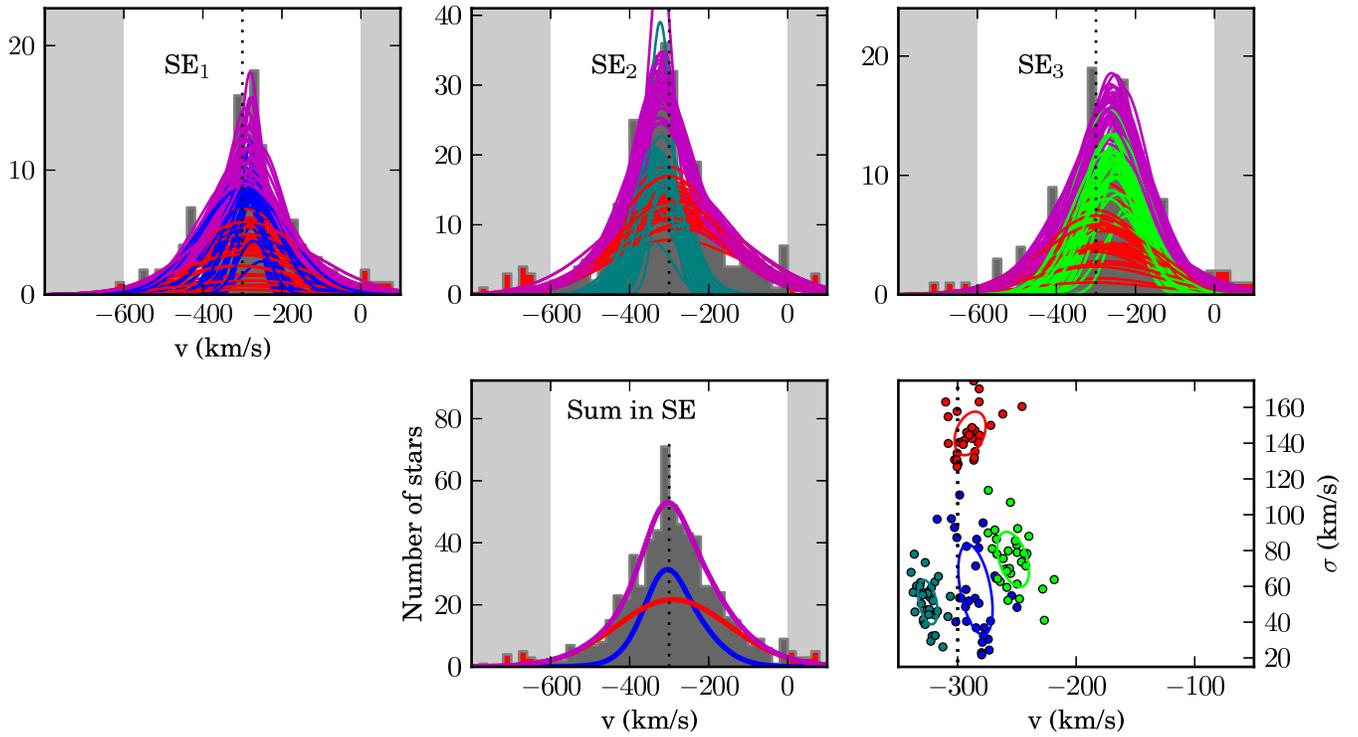}}
\centering
\caption{Same as Figure~\ref{fig_NE1_subs}, but for the SE region. 
 $\rm SE_1$, $\rm SE_2$ and $\rm SE_3$ are the outer, inner south, and inner
  north subregions, respectively.}
\label{fig_SE_subs}
\end{figure*}

We exclude 36 stars from the tails of the velocity distribution in the
SE minor axis region. While this does not skew the results of the MCMC
fits in one direction or another, it does serve to increase the
uncertainty in the final kinematical parameters; this effect can be
seen in the large scatter in the walker positions in the
parameter-space view in Figure~\ref{fig_SE_subs}. 

The minor axis is a saddle point in the velocity field of M31. Here, the
observed rotation velocity changes from less negative than $-300~\rm km~s^{-1}$
on the north side to more negative than $-300~\rm km~s^{-1}$ on the
south side. The mean velocities of the cold component reflect this
transition. The parameter-space view in the final panel of Figure~\ref{fig_SE_subs} shows
that in subregion $SE_1$ (blue), which lies parallel to and just north
of the minor axis, we
measure a mean velocity slightly less negative than the systemic
velocity of M31. In subregions $SE_2$ (turquoise) and $SE_3$ (green),
situated south and north of the minor axis, respectively, we measure
mean velocities more negative and less negative than $-300~\rm
km~s^{-1}$. Finally, we notice that the fraction of spheroid members
is smallest in subregion $SE_1$, which is centered farthest
from the center of the galaxy than the other two subregions. 

\subsection{SSW Region}
Eleven stars fall into the velocity range excluded due to stream
contamination in the SSW region. The
exclusion does not have a significant effect on the mean velocity or
dispersion of the hot component. The chi-squared probability of a good fit
is lower in the SSW region than  in the others, but exclusion of the
possible stream stars improves the probability by a factor of about
15. 

\begin{figure*}[h]
\scalebox{1.0}{\includegraphics[trim = 15 5 55 360, clip =  true,
  width = 1\textwidth]{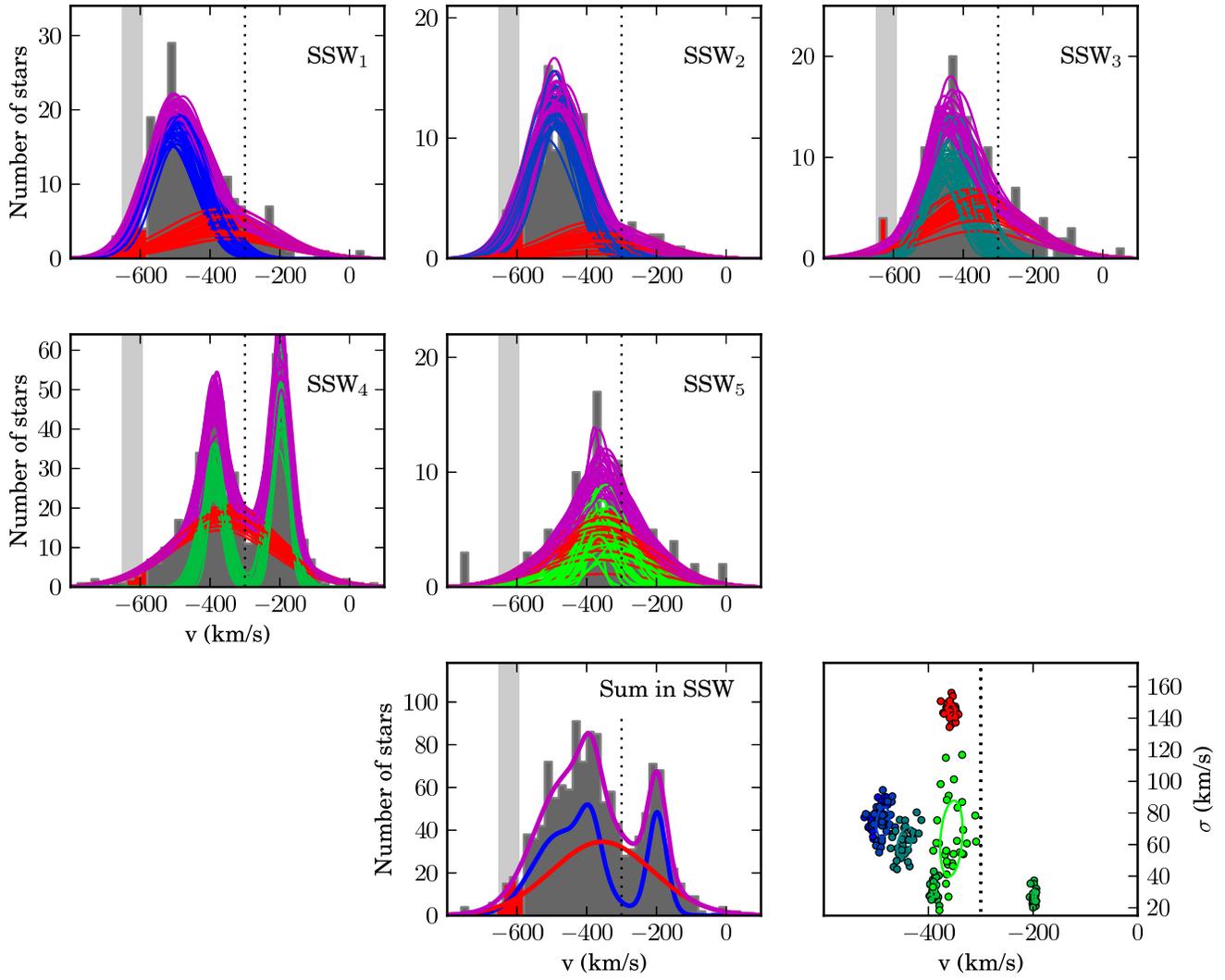}}
\caption{Same as Figure~\ref{fig_NE1_subs}, but for the SSW region.}
\label{fig_SSW_subs}
\end{figure*}

Subregion $\rm SSW_4$ completely contains the galaxy M32, which is treated
as a second cold component. (In a previous trial run, we allowed for the possibility of a
second cold component in each of the five subregions, but the best-fit
fraction of M32 stars in subregions $\rm SSW_{1-3,5}$ was zero.) The
parameter-space view in Figure~\ref{fig_SSW_subs} shows that the mean
velocity and dispersion of M32 are very well constrained around $-200$
and $27~\rm km~s^{-1}$, respectively.

As in the NE regions, the fraction of stars in the spheroid is higher
in the subregions farther from the major axis. 

\clearpage

%FIGURE 13

%FIGURE 14

%FIGURE 15

%FIGURE 16

%FIGURE 17
%\begin{figure}[h]
%\scalebox{0.85}{\includegraphics[trim=90 50 10 20, clip = true]{disk_profiles_nogss.pdf}}
%\caption{Kinematical parameters of the cold component in
%  each subregion, with $1\sigma$ error bars, before exclusion of
%  velocity ranges corresponding to GSS debris. The left panel shows
%  the rotation velocity of each cold component (corrected for offset
%  from the major axis using
%  Equation~\ref{eq_rot}) against the mean $r_{\rm deproj,
%    subregion}$. The three SE subregions have been excluded because
%  they do not yield constraints on the rotation velocity.
% The right panel shows the velocity dispersion. The yellow circle
%  represents M32. Both the velocity and dispersion follow reasonable
%  trends, indicating that our minimal assumption of a locally cold
%  disk adequately accounts for disk contamination.}

%\label{fig_disk_nogss}
%\end{figure}

%FIGURE 17
\begin{figure}[h]
\scalebox{01}{\includegraphics[trim=20 80 10 440, clip = true, width
  = 1\textwidth]{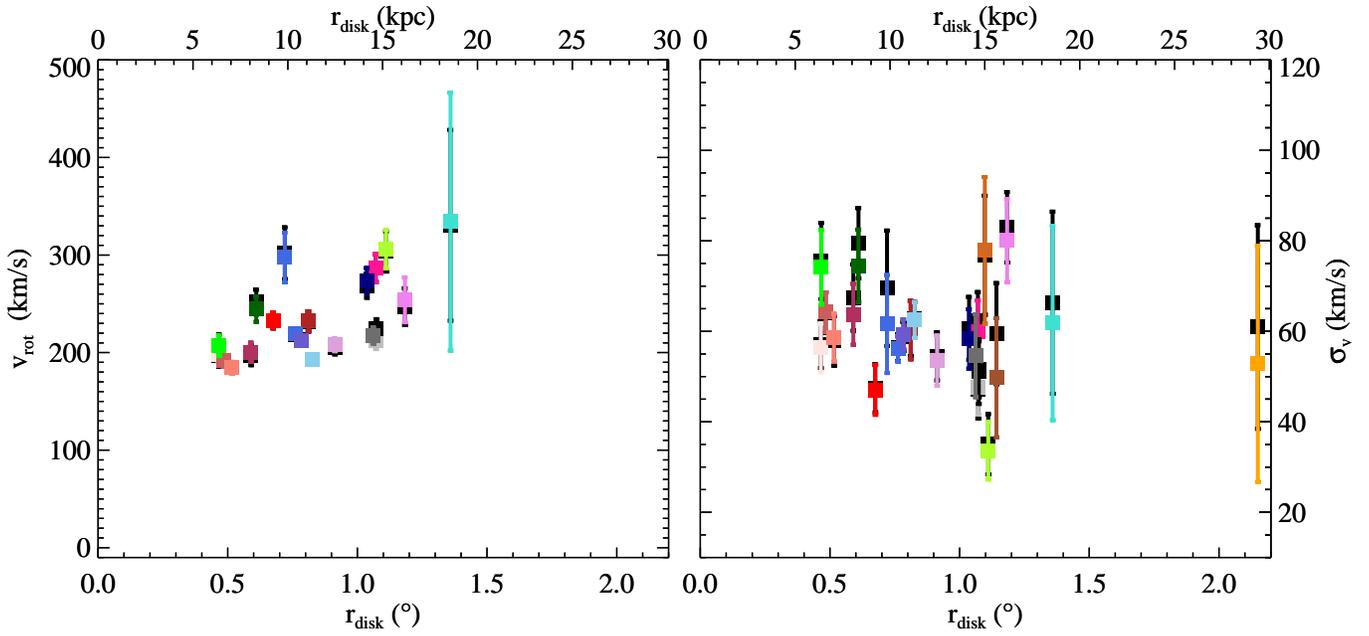}}
\centering
\caption{Kinematical parameters of the cold component in
  each subregion, with $1\sigma$ error bars, before exclusion of
  velocity ranges corresponding to GSS debris (black points) and after
  exclusion (colored points). Note that exclusion of tidal debris has
 a negligible effect on the cold component. These cold components
  correspond to M31's stellar disk. The left panel shows
  the rotation velocity of each cold component (computed using
  Equation~\ref{eq_rot}) vs. the mean $r_{\rm deproj}$ of each
  subregion. The three SE subregions have been excluded because
  they do not yield meaningful constraints on the rotation velocity.
 The right panel shows the velocity dispersion. The velocity and
 dispersion follow the trends expected for M31's stellar disk,
 supporting the idea that that our minimal assumption of a locally cold
  disk adequately accounts for disk contamination. }

\label{fig_disk}
\end{figure}

\begin{deluxetable*}{lllccccccccccc}
\tabletypesize{\scriptsize}
\tablecaption{Kinematical Disk Parameters Corrected for Tidal Debris}
\tablewidth{0 pt}
\tablehead{
\colhead{Subregion}&
\colhead{$\xi~(^\circ)$}&
\colhead{$\eta~(^\circ)$}&
\colhead{$R_{\rm proj}$ (kpc)}&
\colhead{$r_{\rm disk}$ (kpc)}&
\colhead{$N_{\rm region}$}&
\colhead{$\overline{v} - \overline{v}_{M31}~(\rm km~s^{-1})$}&
\colhead{$\sigma_v~(\rm km~s^{-1})$}&
\colhead{$N_{\rm cold}/N_{\rm total}$}&
\colhead{$\chi^2$ prob}&
}
\startdata

\vspace{2mm}
$\rm NE1_1$ & ~~0.284 & ~~0.363 & 6.311 & 6.35 & 220 & $+$192.4 $\pm$ ~5.8
& 56.4 $\pm$ ~5.4 & 0.725 & 0.706\\
\vspace{2mm}
$\rm NE1_2$ & ~~0.287 & ~~0.364 & 6.36 & 6.61 & 227 & $+$186.8 $\pm$ ~6.1
& 64.2 $\pm$ ~4.4 & 0.806  & 0.753\\
\vspace{2mm}
$\rm NE1_3$ & ~~0.286 & ~~0.359 & 6.33 & 7.05 & 236 & $+$179.4 $\pm$ ~6.0
& 58.6 $\pm$ ~5.3 & 0.729  & 0.191\\
\vspace{2mm}
$\rm NE1_4$ & ~~0.321 & ~~0.349 & 6.56 & 8.06 & 234 & $+$184.5 $\pm$ ~8.9
& 63.8 $\pm$ ~6.7 & 0.602  & 0.220\\
\vspace{2mm}
$\rm NE1_5$ & ~~0.362 & ~~0.328 & 6.74 & 9.22 & 164 & $+$181.8 $\pm$ ~6.4
& 47.0 $\pm$ ~5.5 & 0.599  & 0.933\\
\vspace{2mm}
$\rm NE1_6$ & ~~0.394 & ~~0.281 & 6.67 & 11.01 & 252 & $+$137.0 $\pm$ ~6.0
& 60.2 $\pm$ ~6.5 & 0.700  & 0.862\\
\vspace{2mm}
$\rm NE1_7$ & ~~0.422 & ~~0.180  & 6.30 & 14.64 & 282 & $+$103.6 $\pm$ ~5.2
& 60.0 $\pm$ ~6.8 & 0.733  & 0.520\\
\vspace{2mm}
$\rm NE2_1$ & ~~0.463 & ~~0.597 & 10.36 & 10.42 & 306 & $+$213.9 $\pm$ ~3.7
& 56.1 $\pm$ ~2.8 & 0.892  & 0.767\\
\vspace{2mm}
$\rm NE2_2$ & ~~0.459 & ~~0.594 & 10.30 & 10.71  & 264 & $+$206.9 $\pm$
~4.2 & 52.3 $\pm$ ~3.3 & 0.943  & 1.000\\
\vspace{2mm}
$\rm NE2_3$ & ~~0.449 & ~~0.583 & 10.15 & 11.30 & 247 & $+$188.4 $\pm$ ~4.7
& 62.5 $\pm$ ~4.0 & 0.893  & 0.005\\
\vspace{2mm}
$\rm NE2_4$ & ~~0.481 & ~~0.562 & 10.25 & 12.48 & 135 & $+$198.5 $\pm$ ~6.3
& 53.5 $\pm$ ~5.6 & 0.749  & 0.979\\
\vspace{2mm}
$\rm NE2_5$ & ~~0.551 & ~~0.434 & 9.63 & 14.17 & 108 & $+$179.0 $\pm$ ~8.7
& 58.3 $\pm$ ~6.6 & 0.743  & 0.987\\
\vspace{2mm}
$\rm NE2_6$ & ~~0.568 & ~~0.366 & 9.26 & 16.17 & 120 & $+$133.5 $\pm$
12.1 & 80.1 $\pm$ ~9.2 & 0.861 & 1.000\\
\vspace{2mm}
$\rm NE3_1$ & ~~0.653 & ~~0.837 & 14.56 & 14.65 & 339 & $+$206.6 $\pm$ ~8.2
& 47.7 $\pm$ 6.0 & 0.770  & 1.000\\
\vspace{2mm}
$\rm NE3_2$ & ~~0.639 & ~~0.805 & 14.12 & 14.68 & 280 & $+$218.7 $\pm$ ~9.4
& 51.1 $\pm$ ~7.1 & 0.827  & 0.104\\
\vspace{2mm}
$\rm NE3_3$ & ~~0.609 & ~~0.704 & 12.83 & 14.51 & 240 & $+$206.2 $\pm$ ~7.9
& 54.6 $\pm$ ~9.2 & 0.820 & 0.999\\
\vspace{2mm}
$\rm SE_1$ & ~~0.426 & $-$0.241 & 6.68 & 29.38 & 130 & ~$+$16.0 $\pm$ 14.0
& 52.8 $\pm$ 26.1 & 0.448  & 0.424 \\
\vspace{2mm}
$\rm SE_2$ & ~~0.193 & $-$0.170 & 3.62 & 15.61 & 355 & ~$-$25.5 $\pm$ ~8.9
& 49.8 $\pm$ 13.2 & 0.345  & 0.340 \\
\vspace{2mm}
$\rm SE_3$ & ~~0.281 & $-$0.036 & 3.95 & 15.00 & 199 & ~$+$46.3 $\pm$ 14.4
& 77.9 $\pm$ 16.2 & 0.567  & 0.912 \\
\vspace{2mm}
$\rm SSW_1$ & $-$0.262 & $-$0.358 & 6.07 & 6.38 & 254 & $-$200.9 $\pm$
10.1 & 74.2 $\pm$ ~8.2 & 0.647 & 0.489\\
\vspace{2mm}
$\rm SSW_2$ & $-$0.219 & $-$0.425 & 6.54 & 8.34 & 149 & $-$183.7 $\pm$
~9.8 & 74.5 $\pm$ ~8.0 & 0.779  & 0.998 \\
\vspace{2mm}
$\rm SSW_3$ & $-$0.111 & $-$0.371 & 5.31 & 9.84 & 175 & $-$145.3 $\pm$
12.5 & 61.6 $\pm$ 10.8 & 0.449  & 0.433 \\
\vspace{2mm}
$\rm SSW_4$ & $-$0.011 & $-$0.404 & 5.54 & 15.18 & 608 & ~$-$88.5 $\pm$
~5.7 & 33.6 $\pm$ ~6.4 & 0.191  & 0.002 \\
\vspace{4mm}
$\rm SSW_5$ & $-$0.098 & $-$0.366 & 5.19 & 18.58 & 127 & ~$-$54.7 $\pm$
21.6 & 61.8 $\pm$ 21.5 & 0.340  & 5.34 $\times 10^{-9}$ \\
\vspace{2mm}
{$\rm SSW_4 ~(M32)$\tablenotemark{a}} & $-$0.011 & $-$0.404 & 5.54 & 15.18 & 608 & ~$+$102.2 $\pm$
2.7 & 26.4 $\pm$ ~3.2 & 0.270  & 0.002 \\
\tableline \vspace{-3mm}
\enddata
\label{tab_disk}

\tablenotetext{a}{This second cold component in the $\rm SSW_4$
  subregion corresponds to members of the M31 satellite M32. It is
  unrelated to M31's stellar disk and is only included in this table
  for the sake of completeness.}

\end{deluxetable*}

\end{document}